\documentclass[reprint,amsmath,amssymb,amsfonts,aps,floatfix]{revtex4-2}

\usepackage{natbib}
\usepackage{graphicx,color,rotating}
\usepackage[latin1]{inputenc}
\usepackage{textcomp}
\usepackage{dcolumn}
\usepackage{bm}     
\usepackage{upgreek}
\usepackage{subdepth}  			
\usepackage{epstopdf}   		

\usepackage{array}
\newcolumntype{L}[1]{>{\raggedright\let\newline\\\arraybackslash\hspace{0pt}}m{#1}}
\newcolumntype{C}[1]{>{\centering\let\newline\\\arraybackslash\hspace{0pt}}m{#1}}
\newcolumntype{R}[1]{>{\raggedleft\let\newline\\\arraybackslash\hspace{0pt}}m{#1}}

\usepackage{hyperref}					
\usepackage{xcolor}						
\hypersetup{			     			
    colorlinks,
    allcolors={blue}
}
\newcommand{\upspace}{\rule{0ex}{2.5ex}}

\newcommand{\mr}[1]{\ensuremath{\mathrm{#1}}}
\newcommand{\myvec}[1]{\bm{#1}}
\newcommand{\ee}{\mathrm{e}}
\newcommand{\ii}{\mathrm{i}}
\newcommand{\dm}{\mathrm{d}}

\newcommand{\avr}[1]{\big\langle #1 \big\rangle}

\DeclareMathOperator{\re}{Re}

\newcommand{\pp}{\partial}

\newcommand{\nablabf}{\boldsymbol{\nabla}}

\newcommand{\rot}{\nablabf\times}
\renewcommand{\div}{\nablabf\cdot}

\newcommand{\dpst}{\displaystyle}

\newcommand{\eee}{\myvec{e}}
\newcommand{\een}{\myvec{e}}

\newcommand{\FFFrad}{\myvec{F}^\mathrm{rad}}

\newcommand{\FFFradII}{\FFFrad_{11}}
\newcommand{\FFFradin}{\FFFrad_{2,\mr{in}}}

\newcommand{\Imat}{\textbf{\textsf{I}}}

\newcommand{\kc}{k_\mathrm{c}}

\newcommand{\ks}{k_\mathrm{s}}

\newcommand{\nnn}{\myvec{n}}

\newcommand{\rrr}{\myvec{r}}

\newcommand{\rhat}{{\hat{r}}}

\newcommand{\uuu}{\myvec{u}}

\newcommand{\vvv}{\myvec{v}}

\newcommand{\zerovec}{\boldsymbol{0}}

\newcommand{\calO}{\mathcal{O}}

\newcommand{\Bc}{B_\mr{c}}

\newcommand{\Bbc}{B_\mr{c}^\mr{b}}

\newcommand{\cO}{c_0}
\newcommand{\cOO}{c_0^{\infty}}
\newcommand{\cOOsqr}{c_0^{\infty 2}}

\newcommand{\cpO}{c_{p0}}

\newcommand{\cpOOTi}{\tilde{c}^\infty_{p0}}
\newcommand{\cpOOp}{c^{\infty \prime}_{p0}}

\newcommand{\Dth}{D^\mathrm{th}}
\newcommand{\DthO}{D^\mathrm{th}_0}
\newcommand{\DthOO}{D^\mathrm{th \infty}_0}
\newcommand{\DthOOp}{D^\mathrm{th \infty \prime}_0}

\newcommand{\Eac}{E_\mathrm{ac}}

\newcommand{\kthO}{k^\mathrm{th}_0}
\newcommand{\kthOO}{k^\mathrm{th \infty}_0}
\newcommand{\kthOOTi}{\tilde{k}^\mathrm{th \infty}_0}

\newcommand{\kapSOOTi}{\tilde{\kappa}_{s0}^{\infty}}

\newcommand{\kapS}{\kappa_s}
\newcommand{\kapSO}{\kappa_{s0}}
\newcommand{\kapSOO}{\kappa_{s0}^{\infty}}

\newcommand{\kapsO}{\kappa_{s0}}

\newcommand{\PE}{\textit{P\'e}}

\newcommand{\Smn}[2]{S_{#1,#2}}

\newcommand{\DTO}{\Delta T_0}
\newcommand{\DTOsurf}{\Delta T_0^\mr{surf}}

\newcommand{\td}{t_\mr{d}^\mr{th}}

\newcommand{\ain}{\alpha_{i,n}}

\newcommand{\acn}{\alpha_{\mr{c},n}}
\newcommand{\acnsc}{\acn^\mr{sc}}
\newcommand{\acnscO}{\acn^\mr{sc,0}}

\newcommand{\acnp}{\alpha_{\mr{c},n}^\prime}

\newcommand{\asn}{\alpha_{\mr{s},n}}
\newcommand{\asnsc}{\asn^\mr{sc}}

\newcommand{\asnp}{\alpha_{\mr{s},n}^\prime}

\newcommand{\atn}{\alpha_{\mr{t},n}}
\newcommand{\atnsc}{\atn^\mr{sc}}

\newcommand{\atnp}{\alpha_{\mr{t},n}^\prime}

\newcommand{\alphap}{\alpha_p}

\newcommand{\alphapO}{{\alpha_{p0}}}

\newcommand{\delt}{\delta_t}

\newcommand{\dels}{\delta_s}

\newcommand{\etaB}{\eta^\mathrm{b}}
\newcommand{\etaBI}{\eta_1^\mathrm{b}}
\newcommand{\etaBIin}{\eta_1^\mathrm{b,in}}
\newcommand{\etaBO}{\eta^\mathrm{b}_0}
\newcommand{\etaBOO}{\eta^\mathrm{b \infty}_0}

\newcommand{\etaIB}{\eta^\mathrm{b}_1}

\newcommand{\etaO}{\eta_0}
\newcommand{\etaOO}{\eta_0^\infty}
\newcommand{\etaI}{\eta_1}
\newcommand{\etaIin}{\eta_1^\mr{in}}

\newcommand{\etaOTi}{\tilde{\eta}_0}

\newcommand{\Gams}{\Gamma_\mathrm{s}}

\newcommand{\gamO}{\gamma_0}

\newcommand{\nuI}{\nu_1}

\newcommand{\nuBOO}{\nu_0^\mathrm{b \infty}}

\newcommand{\nuIB}{\nu_1^\mathrm{b}}
\newcommand{\nuBI}{\nu_1^\mathrm{b}}
\newcommand{\nuO}{\nu_0}
\newcommand{\nuOO}{\nu_0^\infty}

\newcommand{\surfP}{\partial\Omega_0}

\newcommand{\phiIc}{\phi_\mathrm{1c}}

\newcommand{\phiIcin}{\phi_\mr{1c}^\mr{in}}

\newcommand{\phiIcsc}{\phi_\mr{1c}^\mr{sc}}

\newcommand{\psiIsc}{\psibf_{1}^\mr{sc}}
\newcommand{\psiIp}{\psibf'_{1}}

\newcommand{\psibf}{\bm{\psi}}

\newcommand{\psibfIsc}{\psibf^\mr{sc}_1}
\newcommand{\psibfIp}{\psibf^\mr{\prime}_1}

\newcommand{\cP}{c_p}

\newcommand{\kO}{k_0}

\newcommand{\pIin}{p_1^\mr{in}}
\newcommand{\pIsc}{p_1^\mr{sc}}
\newcommand{\pIscn}{p_{1,n}^\mr{sc}}

\newcommand{\pI}{p_1}

\newcommand{\pIIin}{p_2^\mr{in}}
\newcommand{\pIIsc}{p_2^\mr{sc}}

\newcommand{\TO}{T_0}
\newcommand{\TOO}{T^\infty_0}

\newcommand{\vIr}{v_{1r}}

\newcommand{\vIrp}{v_{1r}^{\prime}}

\newcommand{\vIth}{v_{1\theta}}

\newcommand{\vIthp}{v_{1\theta}^{\prime}}
\newcommand{\vvvIin}{\myvec{v}_1^\mathrm{in}}
\newcommand{\vvvIsc}{\myvec{v}_1^\mathrm{sc}}
\newcommand{\vvvIp}{\myvec{v}'_1}

\newcommand{\vvvI}{\vvv_1}

\newcommand{\vvvII}{\vvv_2}
\newcommand{\vvvIIin}{\vvv_2^\mr{in}}
\newcommand{\vvvIIsc}{\vvv_2^\mr{sc}}

\newcommand{\xO}{x_0}

\newcommand{\phiI}{\phi_1}
\newcommand{\phiIin}{\phi_1^\mathrm{in}}

\newcommand{\phiIsc}{\phi_1^\mathrm{sc}}
\newcommand{\phiIscn}{\phi_{1,n}^\mathrm{sc}}
\newcommand{\phiIp}{\phi'_1}

\newcommand{\rhoO}{\rho_0}
\newcommand{\rhoOO}{\rho_0^{\infty}}
\newcommand{\rhoOOTi}{\tilde{\rho}_0^{\infty}}
\newcommand{\rhoOOp}{\rho_0^{\infty \prime}}

\newcommand{\rhoI}{\rho_1}

\newcommand{\SIMHz}{\textrm{MHz}}

\newcommand{\SIJ}{\textrm{J}}

\newcommand{\SIK}{\textrm{K}}

\newcommand{\SIkg}{\textrm{kg}}
\newcommand{\SIkgm}{\textrm{kg}\:\textrm{m$^{-3}$}}

\newcommand{\SIm}{\textrm{m}}

\newcommand{\SImm}{\textrm{mm}}
\newcommand{\SImum}{\textrm{\textmu{}m}}

\newcommand{\SIMPa}{\textrm{MPa}}

\newcommand{\SImPas}{\textrm{mPa}\:\textrm{s}}

\newcommand{\SIs}{\textrm{s}}
\newcommand{\SIms}{\textrm{ms}}

\newcommand{\SIW}{\textrm{W}}

\newcommand{\SImuW}{\textrm{\textmu{}W}}
\newcommand{\SIGW}{\textrm{GW}}
\newcommand{\nn}{\nonumber}
\newcommand{\beq}[1]{\begin{equation} \eqlab{#1}}
\newcommand{\eeq}{\end{equation}}
\newcommand{\bsub}{\begin{subequations}}
\newcommand{\esub}{\end{subequations}}
\def\bal#1\eal{\begin{align}#1\end{align}}
\def\balat#1#2\ealat{\begin{alignat}{#1} #2 \end{alignat}}
\def\bsubal#1 #2\esubal{\bsuba{#1}\begin{align}#2\end{align} \esuba}     
\def\bsubalat#1#2#3\esubalat{\bsuba{#1} \begin{alignat}{#2} #3 \end{alignat} \esuba}
\newcommand{\bsuba}[1]{\bsub \eqlab{#1}}
\newcommand{\esuba}{\esub}

\newcommand{\eqlab}[1]{\label{eq:#1}}
\renewcommand{\eqref}[1]{Eq.~(\ref{eq:#1})}
\newcommand{\eqnoref}[1]{(\ref{eq:#1})}

\newcommand{\eqsref}[2]{Eqs.~(\ref{eq:#1}) and~(\ref{eq:#2})}
\newcommand{\eqsnoref}[2]{(\ref{eq:#1}) and~(\ref{eq:#2})}
\newcommand{\eqssref}[3]{Eqs.~(\ref{eq:#1}), (\ref{eq:#2}) and~(\ref{eq:#3})}
\newcommand{\eqsssref}[4]{Eqs.~(\ref{eq:#1}), (\ref{eq:#2}), (\ref{eq:#3}) and~(\ref{eq:#4})}

\newcommand{\figref}[1]{Fig.~\ref{fig:#1}}

\newcommand{\figlab}[1]{\label{fig:#1}}
\newcommand{\appref}[1]{Appendix~\ref{sec:#1}}

\newcommand{\appsref}[2]{Appendices~\ref{sec:#1} and~\ref{sec:#2}}

\newcommand{\secref}[1]{Section~\ref{sec:#1}}

\newcommand{\seclab}[1]{\label{sec:#1}}
\newcommand{\tabref}[1]{Table~\ref{tab:#1}}

\newcommand{\tablab}[1]{\label{tab:#1}}

\newcommand{\xth}{x_\mr{D}^\mr{th}}
\newcommand{\xc}{x_\mathrm{c}}

\newcommand{\xs}{x_\mathrm{s}}

\newcommand{\xcp}{x_\mathrm{c}^{\prime}}
\newcommand{\xsp}{x_\mathrm{s}^{\prime}}

\newcommand{\Phiac}{\Phi_\mathrm{ac}}

\newcommand{\sigmabf}{\bm{\sigma}}

\newcommand{\cL}{c_\mathrm{lo}}
\newcommand{\cLO}{c_\mathrm{lo0}}
\newcommand{\cT}{c_\mathrm{tr}}
\newcommand{\cTO}{c_\mathrm{tr0}}

\newcommand{\uuuI}{\myvec{u}_1}

\definecolor{darkgreen}{rgb}{0.00, 0.50, 0.00}
\definecolor{DARKGREEN}{rgb}{0.00, 0.50, 0.00}
\definecolor{RED}{rgb}{1.00, 0.00, 0.00}
\definecolor{GREEN}{rgb}{0.00, 1.00, 0.00}
\definecolor{BLUE}{rgb}{0.00, 0.00, 1.00}
\definecolor{MAGENTA}{rgb}{1.00, 0.00, 1.00}

\newcommand{\hpn}[1]{h_{#1}^{(+)}}
\newcommand{\hmn}[1]{h_{#1}^{(-)}}
\newcommand{\Ipn}[1]{I_{#1}^{(+)}}
\newcommand{\Imn}[1]{I_{#1}^{(-)}}
\newcommand{\hpnp}[1]{h_{#1}^{(+)\prime}}
\newcommand{\hmnp}[1]{h_{#1}^{(-)\prime}}
\newcommand{\hpnpp}[1]{h_{#1}^{(+)\prime\prime}}
\newcommand{\hmnpp}[1]{h_{#1}^{(-)\prime\prime}}

\begin{document}

\title{Acoustic radiation force on a heated spherical particle in a fluid including scattering and microstreaming from a standing ultrasound wave}

\author{Bj\o rn G. Winckelmann}
\email{winckel@dtu.dk}
\affiliation{Department of Physics, Technical University of Denmark,\\
DTU Physics Building 309, DK-2800 Kongens Lyngby, Denmark}

\author{Henrik Bruus}
\email{bruus@fysik.dtu.dk}
\affiliation{Department of Physics, Technical University of Denmark,\\
DTU Physics Building 309, DK-2800 Kongens Lyngby, Denmark}

\date{29 June 2023}

\begin{abstract}
Analytical expressions are derived for the time-averaged, quasi-steady, acoustic radiation force on a heated, spherical, elastic, solid microparticle suspended in a  fluid and located in an axisymmetric incident acoustic wave. The heating is assumed to be spherically symmetric, and the effects of particle vibrations, sound scattering, and acoustic microstreaming are included in the calculations of the acoustic radiation force. It is found that changes in the speed of sound of the fluid due to temperature gradients can significantly change the force on the particle, particularly through perturbations to the microstreaming pattern surrounding the particle. For some fluid-solid combinations, the effects of particle heating even reverse the direction of the force on the particle for a temperature increase at the particle surface as small as 1 K.

\end{abstract}

\maketitle


\section{Introduction}
A particle suspended in a fluid perturbed by an acoustic wave experiences a time-averaged force, termed the acoustic radiation force $\FFFrad$.
Theoretical studies of $\FFFrad$ date back to King in 1934 \cite{King1934} who assumed the particle to be incompressible and the surrounding fluid to be ideal, meaning zero viscosity and zero thermal conductivity. Subsequently, Yosioka and Kawasima \cite{Yosioka1955} included the effects of particle compressibility in 1955, and the results were summarized and expressed on potential form by Gor'kov \cite{Gorkov1962} in 1962.
Doinikov published two series of papers taking into account the effects of fluid viscosity \cite{Doinikov1994a, Doinikov1994} in 1994 and the effects of heat conduction \cite{Doinikov1997, Doinikov1997a, Doinikov1997b} in 1997, where he included both the linear scattering of the acoustic wave and the nonlinear steady acoustic microstreaming developing around the particle.
More recent developments were made by Settnes and Bruus in 2012 \cite{Settnes2012}, Karlsen and Bruus in 2015 \cite{Karlsen2015}, and Doinikov, Fankhauser, and Dual in 2021 \cite{Doinikov2021}.
A detailed study of $\FFFrad$ on small particles in a thermoviscous fluid was conducted in our recent work \cite{Winckelmann2023}. Here, the effects of particle vibrations, acoustic scattering, temperature and density dependent material parameters, and thermoviscous microstreaming were included in the analytical derivation of $\FFFrad$, and it was found that microstreaming effects may dominate $\FFFrad$ when the viscous- and thermal boundary layer widths $\dels$ and $\delt$ are comparable to or larger than the particle radius $a$. The importance of the microstreaming patterns for $\FFFrad$, alongside recent studies of the acoustic body force due to temperature gradients by Joergensen and Bruus \cite{Joergensen2021}, motivated the present study of the acoustic radiation force on a heated spherical microparticle.

In this paper we derive analytical expressions for $\FFFrad$ on a spherical elastic particle in a  Newtonian fluid, including adiabatic acoustic scattering and microstreaming, based on an extension of the theoretical framework presented in Refs.~\cite{Doinikov1994a, Doinikov1994, Doinikov1997, Winckelmann2023}. In this extension we include a quasi-steady background temperature field with gradients, which we calculate from a purely diffusive heat equation. The particle radius $a$ is assumed to be much smaller than the wavelength $\lambda$ of the incident acoustic wave. The acoustic field is assumed to be adiabatic, which is a good approximation in the limit of $5\delt \lesssim a \ll \lambda$ \cite{Winckelmann2023}. Previously, Lee and Wang in 1984 and 1988 \cite{Lee1984, Lee1988} studied the special case of a heated (or cooled) heavy rigid sphere in an ideal inviscid gas, and without taking acoustic microstreaming into account in their analysis. They only considered a short-ranged temperature field, but here we argue that the primary change of $\FFFrad$ is caused by heating of the bulk fluid surrounding the particle, and we therefore reach different conclusions than Lee and Wang.

Unlike previous studies of thermal and viscous contributions to $\FFFrad$ \cite{Doinikov1994a, Doinikov1994, Doinikov1997, Doinikov1997a, Doinikov1997b, Settnes2012, Karlsen2015, Winckelmann2023}, we find that effects of externally generated thermal gradients may alter $\FFFrad$ on particles in the long-wavelength limit even for small boundary layer widths $\delta\ll a$. This may lead to new possibilities for acoustic handling of above $\SImum$-sized particles at MHz ultrasound frequencies through the use of heat sources.

The paper is structured as follows: governing equations are presented in \secref{gov_eq}, our mathematical model is presented and solved in \secref{model}, the results for an incident, standing, plane wave are analyzed in \secref{ResultsPlaneWave}, and finally we conclude in \secref{Conclusion}. Some mathematical details are presented in \appsref{T0_approximation}{S_coeff}, and supporting \textsc{Matlab} scripts, numerical simulations in \textsc{Comsol Multiphysics}, and details on material parameters are provided in the Supplemental Material~\footnote{See Supplemental Material at
\url{https://bruus-lab.dk/files/Winckelmann_Frad_heated_sphere_suppl.zip}
for details on numerical simulations in COMSOL Multiphysics and comments on temperature dependent material parameters.}.\\[-10mm]

\section{Governing equations}
\seclab{gov_eq}
\vspace*{-5mm}
Our model includes an isotropic, elastic, solid particle suspended in a Newtonian fluid. The particle is assumed to be heated by either an external source (e.g.\ a laser) or an internal source (e.g.\  an exothermic chemical reaction), and consequently heat conduction leads to the formation of a temperature gradient in the surrounding fluid. The fluid is perturbed by a monochromatic adiabatic acoustic wave with frequency $f$ and angular frequency $\omega=2\pi f$. All physical fields $g(\rrr,t)$ describing the system in space $\rrr$ and time $t$ are expanded in perturbation series, and the material parameters $q(\rrr,t)$ vary through their dependency of temperature $T$ and density~$\rho$,
 \bsubal{fields_parameters}
 \eqlab{field_pertubation}
 g(\rrr,t)&=g_0(\rrr,t)+\re\Big[g_1(\rrr,t)\,\ee^{-\ii\omega t}\Big] + g_2(\rrr,t),
 \\
 \eqlab{parameter_T_dependent}
 q &= q_0[T_0(\rrr,t)] + \re\Big[q_1(\rrr,t)\,\ee^{-\ii\omega t}\Big],
 \\
 \eqlab{q1_definition}
 q_1(\rrr,t) &= \bigg(\frac{\pp q}{\pp T}\bigg)_{T=T_0} T_1(\rrr,t)+\bigg(\frac{\pp q}{\pp \rho}\bigg)_{\rho=\rho_0}\rho_1(\rrr,t).
 \esubal
The zeroth-order fields $g_0(\rrr,t)$ describe a quiescent fluid, with a background temperature field $T_0(\rrr,t)$, and the zeroth-order parameters $q_0(\rrr,t)$ are assumed to be functions of $T_0$ only, so their density dependency is neglected in the following. The complex-valued first-order fields $g_1$ describe the linear acoustic response, which follows the actuation frequency $f$. The second-order fields $g_2(\rrr,t)$ describe a non-linear response containing small second-order harmonics and a steady time-averaged response. Only the time-averaged second-order effects are considered here, and they are denoted by angled brackets, e.g.\ $\avr{g_2(\rrr,t)}$. The acoustic oscillations and streaming generally depend on the background temperature field $T_0(\rrr,t)$ due to the temperature dependencies of all physical parameters $q_0$. We assume that the temperature field develops on a timescale much slower than an acoustic oscillation period $f^{-1}$, and the complex-valued acoustic fields $g_1(\rrr,t)$ are calculated as steady fields at any given time $t$ using the instantaneous temperature field $T_0(\rrr,t)$.

The objective is to compute $\FFFrad$ by the time average of the stress $\sigmabf$ integrated over the vibrating particle surface $\pp\Omega(t)$ with normal vector $\nnn$,
 \beq{Frad_fundamental}
 \FFFrad =\bigg\langle \oint_{\pp\Omega(t)} \sigmabf\cdot \nnn \: \dm S\bigg\rangle.
 \eeq
Assuming that the particle drift is negligible during an acoustic period, $\FFFrad$ can be written as \cite{Doinikov1994a, Karlsen2015},
 \beq{Frad}
 \FFFrad = \oint_{\surfP} \langle \sigmabf_2-\rho_0\vvvI\vvvI \rangle \cdot \nnn \,\dm S,
 \eeq
where $\surfP$ is the equilibrium surface of the particle, and $\vvv$ is the velocity field of the fluid surrounding the particle.

\subsection{Thermal diffusion}
The temperature field is treated as a transient background field. Heat diffusion is assumed to dominate over the heat convection caused by the acoustic streaming $\avr{\vvvII}$, thus we assume a low P\'{e}clet number,
 \bal\eqlab{Peclet_Number}
 \PE = \frac{L |\avr{\vvv_2}|}{\Dth_0} \ll 1 , \qquad \Dth_0 = \frac{\kthO}{\rhoO  \cpO}.
 \eal
Here, $L$ is the characteristic length scale for heat diffusion, $\big|\avr{\vvv_2}\big|$ is the magnitude of the steady streaming field in the fluid, $\Dth_0$ is the thermal diffusivity, $\rhoO$ is the mass density of the quiescent medium at a given temperature $T_0$, $\kthO$ is the thermal conductivity, and $\cpO$ is the specific heat capacity at constant pressure. The time average of the particle motion is assumed to be zero, and the temperature in both the solid particle and the fluid thus follow the heat diffusion equation,
 \bal
 \eqlab{Heat_diffusion}
 \pp_t T_0 = \frac{1}{\rhoO\cpO}\div\big(\kthO \nablabf T_0\big) +\frac{1}{\rhoO  \cpO} P,
 \eal
where $P$ is the power density absorbed by the medium due to an external or internal source of energy.

We assume that gradients in the temperature field $T_0$ are small enough that material parameters $q_0$ only deviate slightly from their ambient value $q_0^\infty$, and that the change is linear in the deviation $\Delta \TO$ from the ambient temperature,
 \bsubal{T0_assumptions}
 \eqlab{DelT0}
 &\DTO(\rrr,t) = \TO(\rrr,t)-\TOO
 \\
 \eqlab{q0_definition}
 &q_0 \approx q_0^\infty(1+a_q \Delta T_0), \qquad  a_q = \frac{1}{q_0^\infty} \Big(\frac{\pp  q_0}{\pp T}\Big)_{T_0^\infty},
 \\
 \eqlab{Small_thermal_parameter}
 &|a_q \Delta T_0| \ll 1,
 \\
 \eqlab{Derivative_assumption}
 &\text{and terms containing $\pp_rT_0(\rrr,t)$ are neglected}.
 \esubal
The assumption~\eqnoref{Derivative_assumption} is justified by a numerical study presented in the Supplemental Material \cite{Note1}. The assumption is based on the fact that the temperature field $\Delta T_0$ falls off as $\Delta T_0\sim r^{-1}$, where $r$ is the distance to the particle center, whereas $\pp_r T_0\sim r^{-2}$. We argue that the primary perturbation to $\FFFrad$ is due to the enhancement of a directional microstreaming caused by temperature-induced, long-ranged perturbations of the acoustic fields in the bulk. Consequently, the dominating terms are likely to be the ones decaying slowly with $r^{-1}$.

\subsection{Viscous fluid dynamics}
The surrounding fluid is described by the fluid velocity field $\vvv$ and the fluid stress tensor $\sigmabf$ expressed in terms of the dynamic viscosity $\eta$, the bulk viscosity $\etaB$, and the fluid pressure $p$,
 \bal
 \eqlab{sigma_fl}
 \sigmabf = \eta \Big[\nablabf \vvv +(\nablabf \vvv)^\textsf{T}\Big] +
 \big[(\etaB-\tfrac23 \eta)(\div \vvv) - p \big]\Imat .
 \eal
The physical fields are governed by local conservation of mass and momentum. To derive the acoustic equations, we assume the adiabatic condition $\dm s = 0$ on the entropy $s$ per unit mass and apply an equation of state relating $p$ and $\rho$,
 \bsubal{fluid_governing}
 \eqlab{continuity_eq}
 \pp_t\rho &=\div (-\rho \vvv),
 \\
 \eqlab{NS_eq}
 \pp_t(\rho\vvv)&=\div (\sigmabf -\rho \vvv\vvv),
 \\
 \eqlab{Adiabatic}
 \dm s & = \frac{\cP}{T} \,\dm T -\frac{\alphap}{\rho} \, \dm p = 0,
 \\
 \eqlab{Equation_of_state}
 p&=p(\rho).
 \esubal
We also introduce the isentropic compressibility $\kapS$,
 \bal
 \eqlab{compressibility_fl}
 \kapS = \frac{1}{\rho}\Big(\frac{\pp \rho}{\pp p}\Big)_s = \frac{1}{\rho c^2}, \qquad \textrm{(fluids)},
 \eal
where $c$ is the speed of sound in the fluid.

The fluid fields and parameters are expanded as described by \eqref{fields_parameters}, and we assume that the initial state of the fluid is quiscent, $\vvv_0=\zerovec$. With the assumption~\eqnoref{Derivative_assumption}, \eqsref{sigma_fl}{fluid_governing} give the first-order acoustic response,
 \bsubal{fluid_first_order}
 \eqlab{continuity_eq_first_order}
 \ii\omega \frac{1}{c_0^2}p_1 &=\rhoO \div  \vvv_1,
 \\
 \eqlab{NS_eq_first_order}
 -\ii\omega \rho_0 \vvv_1 &= \etaO \nabla^2 \vvv_1 + \big(\etaB_0+\tfrac13  \eta_0\big) \nablabf(\div \vvv_1) - \nablabf p_1  .
 \esubal
The adiabatic assumption in \eqref{Adiabatic} further dictates that the acoustic temperature field $T_1$ is proportional to the pressure $p_1$,
 \beq{T1_def}
 T_1 = \frac{\kapsO (\gamO-1)}{\alphapO} p_1, \quad \gamO = 1+\frac{\alpha_{p0}^2 T_0}{\rhoO \cpO \kapsO},
 \eeq
where we have introduced the usual ratio $\gamO =\cpO/c_{v0}$ of the specific heat capacities.
We then use $\rhoI=\cO^{-2} p_1$ and combine \eqsref{q1_definition}{T1_def},
 \bsubal{eta1_etab1}
 \eqlab{eta1}
 \etaI &= -\Bc \frac{\nuO}{\cO^2} p_1 ,
 \\
 \eqlab{etab1}
 \etaIB &= -\Bbc \frac{\nuO}{\cO^2} p_1 ,
 \\
 \eqlab{Bc}
 \Bc &=\bigg[\frac{1-\gamO}{\alphapO \etaO} \bigg(\frac{\pp \eta}{\pp T}\bigg)_{T_0}
 - \frac{\rhoO}{\etaO}\bigg(\frac{\pp \eta}{\pp  \rho}\bigg)_{\rho_0}\bigg] ,
 \\
 \eqlab{Bt}
 \Bbc &=\bigg[\frac{1-\gamO}{\alphapO \etaO} \bigg(\frac{\pp \etaB}{\pp T}\bigg)_{T_0}
 - \frac{\rhoO}{\etaO}\bigg(\frac{\pp \etaB}{\pp  \rho}\bigg)_{\rho_0}\bigg]  .
 \esubal

The time-averaged second-order terms of \eqsref{sigma_fl}{fluid_governing} describe the non-linear streaming response,
 \bsubal{fluid_second_order}
 \eqlab{continuity_eq_second_order}
 &0 =\div \Big\langle \rho_0 \vvv_2+\frac{1}{c_0^2}p_1\vvv_1 \Big\rangle ,
 \\
 \eqlab{NS_eq_second_order}
 &\zerovec =\div \avr{\sigmabf_2-\rho_0\vvv_1\vvv_1} ,
 \\
 \eqlab{sigma_second_order}
 &\avr{\sigmabf_2}=\eta_0 \Big[\nablabf \avr{\vvv_2} \!+\! \big(\nablabf \avr{\vvv_2}\big)^\textsf{T}\Big] \!+ \! \big(\etaB_0 \!-\! \tfrac{2}{3}\eta_0\big)\big(\div \avr{\vvv_2}\big) \,\Imat
 \nn\\
 &\, -\avr{p_2} \, \Imat + \Big\langle\eta_1 \Big[\nablabf \vvv_1 \!+\! (\nablabf \vvv_1)^\textsf{T}\Big] \!+\! \big(\etaB_1 \!-\! \tfrac{2}{3}\eta_1\big)(\div \vvv_1)\,\Imat \Big\rangle.
 \esubal
We note that the zeroth-order quantities $c_0$, $\rho_0$, $\eta_0$, and $\etaB_0$ depend implicitly on the spatial coordinates $\rrr$ through gradients in the background temperature $T_0(\rrr,t)$.

\subsection{Isotropic elastic solid mechanics}
The linear elastic solid is described by the the mechanical displacement field $\uuu$, and the solid stress tensor $\sigmabf$, which is expressed in terms of the transverse and longitudinal speeds of sound $\cT$ and $\cL$,
 \bal
 \eqlab{sigma_sl}
 \sigmabf = \rho \cT^2 [\nablabf \uuu +(\nablabf \uuu)^\textsf{T}]+\rho(\cL^2-2\cT^2) (\div  \uuu)\,\Imat.
 \eal
The mechanical displacement field $\uuu$ can then be determined in time and space from the Cauchy equation,
 \bal
 \eqlab{N2_sl}
 \rho \pp_t^2\uuu = \div \sigmabf.
 \eal
For solids, the isentropic compressibility $\kapS$ can be expressed as,
 \bal
 \eqlab{compressibility_sl}
 \kapS  = \frac{1}{\rho (\cL^2 - \tfrac43 \cT^2)}, \qquad \textrm{(solids)}.
 \eal

Expanding the fields into a perturbation series with $\uuu_0=\zerovec$, one can describe the acoustic vibrations by,
 \beq{N2_sl_first_order}
 -\rhoO \omega^2 \uuuI = \rhoO \cTO^2 \nabla^2 \uuuI +\rhoO \Big( \cLO^2-\cTO^2 \Big) \nablabf(\div \uuuI),
 \eeq
where we have used assumption \eqnoref{Derivative_assumption}. Following Ref.~\cite{Karlsen2015}, we also define the velocity field $\vvv_1$ and the complex-valued "viscosity" $\etaO$ of the solid by,
 \beq{solid_definitions}
 \vvvI = -\ii\omega \uuu_1
 \quad \text{and} \quad
 \etaO = \ii \frac{\rhoO \cTO^2}{\omega}, \quad \text{for solids}.
 \eeq
The second-order response is not calculated for solids, as the time-averaged second-order velocity field is zero, and the steady thermal expansion is negligible.

\section{Model}
\seclab{model}

\begin{figure*}[t]
\centering
\includegraphics[width=\linewidth]{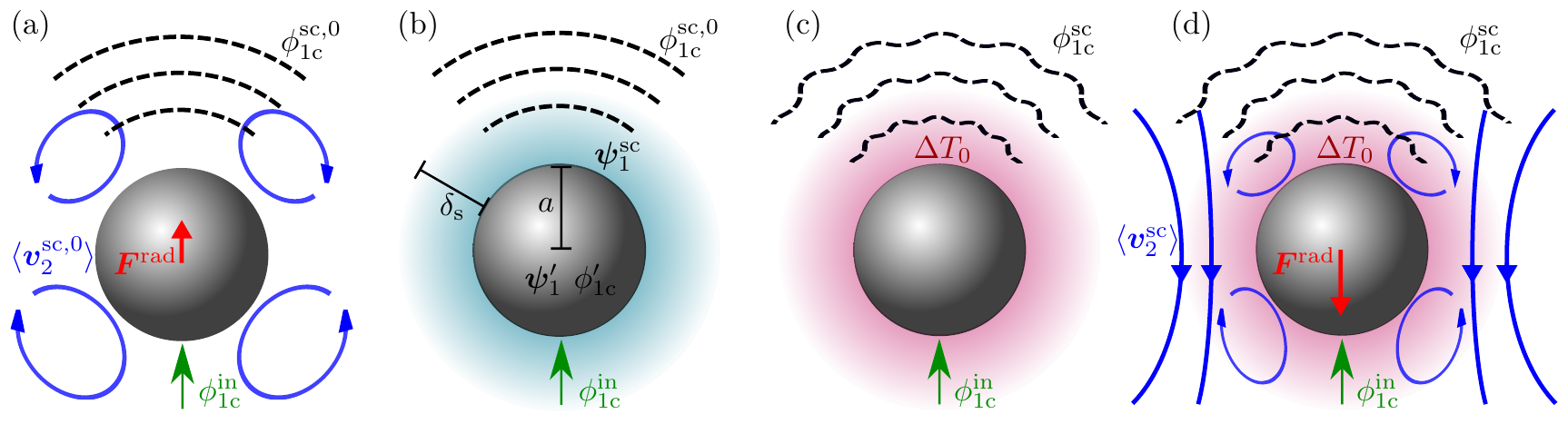}
\caption{\figlab{problem_sketch}
(a) The standard constant-temperature case  $\Delta T_0 = 0$ \cite{Doinikov1994a} of a spherical particle in an incident acoustic wave $\phiIcin$ (green arrow), which gives rise to the acoustic radiation force $\FFFrad$ (red arrow) on the particle through scattered waves (such as $\phiIc^\mr{sc,0}$, dashed lines) and microstreaming $\avr{\vvvII^\mr{sc,0}}$ (quadrupolar-like blue arrows).
(b) Emphasizing further details of the standards case: the viscous scattering $\psibfIsc$ in the boundary layer (light blue) of width $\dels$ \textit{outside} and the transmitted waves $\phiIc'$ and $\psibfIp$ \textit{inside} the particle of radius $a$.
(c) Heating of the particle gives rise to the temperature deviation $\Delta T_0$ (light red) in the surrounding fluid. The scattered acoustic wave $\phiIcsc$ (warped dashed lines) for $\Delta T_0 > 0$ is significantly changed compared to $\phiIc^\mr{sc,0}$ in (a) and (b).
(d) Acoustic scattering on the heated sphere with thermal contributions to both scattering and microstreaming $\avr{\vvvIIsc}$ (blue), now with a significant directional component that leads to a modified acoustic radiation force $\FFFrad$ (red arrow).
}
\end{figure*}

As illustrated in \figref{problem_sketch}, the physical model consists of a spherical, heated, solid particle of radius $a$, which is centered at the origin of a spherical coordinate system $(r,\theta,\varphi)$. The particle is surrounded by a viscous fluid of infinite extent, and an incident pressure wave with axisymmetry around the $z$-axis propagates in the fluid and scatters on the sphere. Since many of the same physical quantities are defined in both the solid and the fluid, we denote all fields and parameters in the solid at $r<a$ with a prime, e.g.\ $\rho_0'$ and $\vvv_1'$, whereas the fluid fields and parameters at $r>a$ remain unprimed, e.g.\ $\rho_0$ and $\vvv_1$. Ratios of parameters of the solid particle relative to the surrounding fluid are denoted by a tilde and the  normalized radial coordinate by a hat,
 \beq{tilde_notation}
 \tilde{q}_0 = \frac{q_0'}{q_0}, \qquad \rhat = \frac{r}{a}.
 \eeq
In the following derivation we largely use the same notation and basic partial-wave expansion as in our previous work~\cite{Winckelmann2023}.

\subsection{The zeroth-order heat diffusion}
\seclab{zeroth_order}

When calculating the background temperature field, we assume that only the spherical solid particle is heated by the power source. Further, it is assumed that the power is uniformly absorbed throughout the particle, so the power density $P$ is given by,
 \bal
 \eqlab{P_form}
 P(\rrr,t)= P_0 \, \Theta(1-\rhat) \Theta(t) ,
 \eal
where $\Theta(\xi)$ is the Heaviside step function.
The assumptions \eqnoref{T0_assumptions} are applied to \eqref{Heat_diffusion} neglecting terms $\big| a_q \Delta T_0 \big| \ll 1$, and we solve for the temperature field $T_0'$ inside the particle ($\rhat<1$), and $T_0$ outside ($\rhat>1$),
 \bsubal{Heat_diffusion_ana}
 \eqlab{Heat_solid_ana}
 \pp_t T_0' &= \DthOOp \nabla^2 T_0' +\frac{1}{\rhoOOp \cpOOp}P_0,  \quad t>0 ,
 \\
 \eqlab{Heat_fluid_ana}
 \pp_t T_0 &= \DthOO \nabla^2 T_0.
 \esubal
Initially for $t<0$, the particle is assumed to be at ambient temperature $T_0^\infty$. Then, at time $t=0$, the heating $P$ is turned on. We apply continuity of temperature and heat flux, and the temperature is held at $T_0^\infty$ infinitely far from the sphere, so the boundary conditions are,
 \bsubal{BC_T0}
 \eqlab{T0_init}
 T_0(r,0)&=T_0'(r,0)=T_0^\infty ,
 \\
 \eqlab{T0_cont}
 T_0(a,t)&=T_0'(a,t) ,
 \\
 \eqlab{dT0_cont}
 \kthOO \pp_r T_0(a,t) &= k_0^{\mr{th}\infty\prime} \pp_r T_0'(a,t) ,
 \\
 \eqlab{T0_finite}
 T_0(\infty,t) &= T_0^\infty .
 \esubal
The heat diffusion problem in \eqsref{Heat_diffusion_ana}{BC_T0} is treated in Ref.~\cite{Goldenberg1952}, where the solution is written as an integral to be numerically evaluated. However, as described in \appref{T0_approximation}, a good analytic approximation can be found for times $t \gtrsim 5\td$, where $\td$ is the characteristic timescale for heat diffusion over a particle radius. For polystyrene at room temperature we obtain,
 \beq{Th_diff_timescale}
 t \gtrsim 5\td = 5 \frac{a^2}{\DthOOp}
 \approx \Big(\frac{a}{5~\SImum}\Big)^2\times 1~\SIms,
 \eeq
which, for the particle sizes we consider here, is well below the timescale for the bulk fluid dynamics. In this long-time limit, the solution for the deviation $\Delta T_0(\rhat,t)$ in the temperature of the fluid from the ambient $T_0^\infty$ becomes,
 \bsubal{T0_ana}
 \Delta T_0(\rhat,t) &=  \frac{\DTOsurf}{\rhat} \, \mr{erfc}\big[\xth(t)\: \rhat\big],
 \;\;
 \text{for }\; t \gtrsim 5\td,
 \\
 \eqlab{DTOsurf_def}
 \xth(t) & = a\:\big(4\DthOO t\big)^{-\frac12}, \quad
 \DTOsurf=\frac{P_0 a^2}{3 \kthOO},
 \esubal
where $\DTOsurf$ is the asymptotic value of the surface temperature of the particle.
In the following theoretical derivation, $\Delta T_0(\rhat,t)$ is the temperature profile~\eqnoref{T0_ana}.

\subsection{The first-order acoustic scattering}
\seclab{first_order}

\vspace*{-2mm} \noindent
The fluid is mechanically perturbed by an external, incident, acoustic wave that scatters on the particle. The first-order fluid fields are split into an incident (in) and a scattered (sc) field,
 \beq{fluid_field_decomposition}
 p_1 = \pIin+\pIsc ,
 \qquad
 \vvv_1 = \vvvIin+\vvvIsc.
 \eeq
Note that $\pIin$ and $\vvvIin$ are the fields in the fluid at \textit{constant ambient temperature} $\DTO = 0$ and \textit{without} the particle,
 \bsuba{Incident_equations_first_order}
 \bal
 \eqlab{continuity_eq_first_order_incident}
 \ii\omega \frac{1}{\cOOsqr} \pIin &=\rhoOO \div  \vvvIin,
 \eal
 \bal
 \eqlab{NS_eq_first_order_incident}
 -\ii\omega \rhoOO \vvvIin &= - \nablabf \pIin + \etaOO \nabla^2 \vvvIin
 \nn \\
 &\quad + \big(\etaBOO+\tfrac13  \etaOO\big) \nablabf(\div \vvvIin)   .
 \eal
 \esuba
The total fields $p_1$ and $\vvvI$ obey the governing equations in the heated fluid with $\DTO > 0$, and therefore the scattered fields $\pIsc$ and $\vvvIsc$ represent the presence of both the particle and of the temperature deviation in the fluid. Subtracting \eqref{Incident_equations_first_order} from \eqref{fluid_first_order}, we find the set of equations describing the scattered fields as,
 \bsuba{Scattered_equations}
 \bal
 \eqlab{Scattered_cont}
 &\bigg(\frac{\ii\omega}{\cO^2} - \frac{\ii\omega}{\cOOsqr} \bigg)\pIin +   \frac{\ii\omega}{\cO^2}\pIsc = (\rhoO \!-\! \rhoOO)\div\vvvIin \!+\! \rhoO \div\vvvIsc,
 \\
 \eqlab{Scattered_NS}
 &-\ii\omega (\rhoO-\rhoOO)\vvvIin -\ii\omega \rhoO \vvvIsc =
 \nn \\
 & (\etaO \!-\! \etaOO) \nabla^2 \vvvIin \!+\! \Big[\big(\etaBO \!+\! \tfrac13 \etaO\big) \!-\!\big(\etaBOO \!+\! \tfrac13 \etaOO \big)\Big]\nablabf(\div \vvvIin)
 \nn\\
 & +\etaO \nabla^2 \vvvIsc +(\etaBO + \tfrac13 \etaO)\nablabf(\div  \vvvIsc)-\nablabf\pIsc .
 \eal
 \esuba
We note that terms with prefactors $q_0 - q^\infty_0$ are explicitly caused by the gradient in $T_0$, and thus do not appear in our previous work~\cite{Winckelmann2023}. Consequently, in the following we encounter inhomogeneous Helmholtz equations, and we must re-derive the first-order solutions of Ref.~\cite{Winckelmann2023}.

As in Ref.~\cite{Winckelmann2023}, we solve the first-order acoustic scattering problem by Helmholtz decompositions of the velocities $\vvvIin$ (purely compressional) and $\vvvIsc$ of the fluid as well as $\vvvIp$ of the particle,
 \bsubal{Helmholtz_decomposition}
 \eqlab{Helmholtz_decomposition_vIin}
 \vvvIin &= \nablabf \phiIin,
 \\
 \eqlab{Helmholtz_decomposition_vIsc}
 \vvvIsc &= \nablabf \phiIsc + \rot\psiIsc,
 \\
 \eqlab{Helmholtz_decomposition_vIp}
 \vvvIp &= \nablabf \phiIp + \rot\psiIp.
 \esubal

In Ref.~\cite{Winckelmann2023}, we split the scalar potentials $\phiI$ and $\phiIp$ into a compressional long-range part describing the weakly damped bulk waves and a thermal part describing the short-range thermal boundary layer. The thermal boundary layer drops out due to the adiabaticity assumption~\eqnoref{Adiabatic}, so $\phiI$ only refers to the compressional part. Inserting \eqref{Helmholtz_decomposition_vIin} in \eqref{Incident_equations_first_order}, \eqref{Helmholtz_decomposition_vIsc} in \eqref{Scattered_equations}, and \eqsref{Helmholtz_decomposition_vIp}{solid_definitions} in \eqref{N2_sl_first_order}, we derive that
 \bsubal{Helmholtz_equations}
 \eqlab{Helmholtz_phiIp}
 \nabla^2 \phiIp + \kc^{\prime 2} \phiIp &= 0, \quad \kc' = \frac{\omega}{\cLO'},
 \\
 \eqlab{Helmholtz_psiIp}
 \nabla^2 \psiIp + \ks^{\prime 2} \psiIp &= 0, \quad \ks' = \frac{\omega}{\cTO'},
 \\
 \eqlab{Helmholtz_psiIsc}
 \nabla^2 \psiIsc + \ks^2 \psiIsc &= 0 , \;\; \ks = \frac{1\!+\!\ii}{\dels},
 \!\quad \dels = \Big(\frac{2\etaOO}{\rhoOO\omega}\Big)^\frac12\!,
 \\
 \eqlab{Helmholtz_phiIin}
 \nabla^2 \phiIin + \kO^2 \phiIin &= 0 , \quad \kO = \frac{\omega}{\cOO} ,
 \\
 \eqlab{Helmholtz_pIsc}
 \nabla^2 \pIsc + \kO^2 \pIsc &= \kO^2 2a_c \Delta T_0  \pIin ,
 \\
 \eqlab{vIin}
 & \hspace{-18mm} \vvvIin = \frac{1}{\ii\omega\rhoOO} \nablabf \pIin , \qquad \pIin = \ii \omega \rhoOO \phiIin,
 \\
 \eqlab{vIsc}
 & \hspace{-18mm} \vvvIsc = \frac{1}{\ii\omega\rhoOO} \nablabf \pIsc - \frac{a_\rho \Delta T_0}{\ii\omega\rhoOO} \nablabf \pIin + \rot\psiIsc,
 \\
 \eqlab{ac_arho}
 & \hspace{-18mm}
 a_c = \frac{1}{\cOO}\:\Big(\frac{\pp \cO}{\partial T}\Big)_{\TOO},
 \quad
 a_\rho = \frac{1}{\rhoOO}\:\Big(\frac{\pp \rhoO}{\partial T}\Big)_{\TOO}.
 \esubal
Here, we have introduced the undamped compressional wave numbers $\kO$ and $\kc'$, the shear wave numbers $\ks$ and $\ks'$, and the viscous boundary-layer thickness $\dels$. Only the real part $\kO$ is taken into account in the wave number of the compressional wave in the fluid, as the imaginary part associated with damping is smaller by a factor $\Gamma_c =(\etaBOO+\tfrac43 \etaOO)\omega \kapSOO \ll 1$. We have also neglected factors  $\big| a_q \Delta T_0 \big| \ll 1$ in the results of \eqref{Helmholtz_equations}, except that terms of the form $a_q\Delta T_0 \pIin$ are kept because $|\pIin|\gg |\pIsc|$. For our analysis, we introduce the normalized wave numbers,
 \beq{normalized_wave_numbers}
 \xO = \kO a, \quad\;
 \xs = \ks a, \quad\;
 \xc' = \kc' a, \quad\;
 \xs' = \ks' a,
 \eeq
and we note that we will be working in the long-wavelength limit characterized by the small parameter,
 \beq{xO}
 \xO  \ll 1.
 \eeq

\eqsssref{Helmholtz_phiIp}{Helmholtz_psiIp}{Helmholtz_psiIsc}{Helmholtz_phiIin} are all homogeneous Helmholtz equations with axisymmetric solutions expressed in terms of spherical Bessel functions $j_n(x)$, spherical outgoing Hankel functions $\hpn n(x)$ (called $h_n(x)$ in Ref.~\cite{Winckelmann2023}), and Legendre polynomials $P_n(\cos\theta)$,
\\[-5mm]
 \bsubal{standard_Helmholtz_solutions}
 \eqlab{phiIin}
 \phiIin &= \sum_{n=0}^\infty A_n j_n(\xO \rhat) P_n(\cos\theta),
 \\
 \eqlab{phiIp}
 \phiIp &= \sum_{n=0}^\infty A_n \acnp j_n(\xcp \rhat) P_n(\cos\theta),
 \\
 \eqlab{psiIp}
 \psiIp &= \eee_\varphi \sum_{n=1}^\infty A_n \asnp j_n(\xsp \rhat) \pp_\theta P_n(\cos\theta),
 \\
 \eqlab{psiIsc}
 \psiIsc &= \eee_\varphi \sum_{n=1}^\infty A_n \asnsc \hpn n(\xs \rhat) \pp_\theta P_n(\cos\theta).
 \esubal
The inhomogenous Helmholtz equation~\eqnoref{Helmholtz_pIsc} is solved by a partial-wave expansion for $\pIsc$,
\\[-5mm]
 \bsubal{phiIsc_expansion_equation}
 \eqlab{pIsc_expansion}
 &\pIsc = \sum_{n=0}^\infty \pIscn(\rhat) P_n(\cos\theta),
 \\
 \eqlab{phiIsc_equation}
 & \mathcal{\hat{D}}_\rhat \pIscn(\rhat) =\ii\rhoOO\omega \, 2a_c \Delta T_0 \xO^2 \rhat^2 j_n(\xO \rhat),
 \\
 \eqlab{D_operator}
 & \mathcal{\hat{D}}_\rhat = \bigg[\frac{\dm}{\dm \rhat}\Big(\rhat^2 \frac{\dm}{\dm \rhat} \Big)  - n(n+1)  + \xO^2\rhat^2 \bigg].
 \esubal
To construct a Green's function $G_n(\rhat,\xi)$ for \eqref{phiIsc_equation} that solves $\mathcal{\hat{D}}_\rhat G_n(\rhat,\xi) =\delta(\rhat-\xi)$, we introduce the incoming spherical Hankel functions $\hmn n(x)$. The Green's function $G_n(\rhat,\xi)$, which obeys the conditions of continuity, $G_n(\rhat=\xi^+,\xi)=G_n(\rhat=\xi^-,\xi)$, and the derivative jump, $\pp_\rhat G_n(\rhat=\xi^+,\xi)-\pp_\rhat G_n(\rhat=\xi^-,\xi) = \xi^{-2}$, is,
 \bal\eqlab{Greens_function_phiIscn}
 G_n(\rhat,\xi) &= \Big\{\Big[B_n(\xi)+\frac{\xO}{2\ii}\hmn n (\xO\xi)\Big] \hpn{n} (\xO \rhat)
 \nn\\
 &\qquad + C_n(\xi) \hmn{n}(\xO \rhat) \Big\} 	\Theta(\rhat-\xi)
 \nn\\
 &\quad + \Big\{\Big[C_n(\xi)+\frac{\xO}{2\ii}\hpn n (\xO\xi)\Big] \hmn n(\xO \rhat)
 \nn\\
 &\qquad +B_n(\xi)\hpn n(\xO \rhat)\Big\} \Theta(\xi-\rhat) .
 \eal
Here,  $B_n(\xi)$ and $C_n(\xi)$ are found from the remaining boundary conditions. Then, with $G_n(\rhat,\xi)$ given by \eqref{Greens_function_phiIscn}, the solution to \eqref{phiIsc_equation} that satisfies the Sommerfeld radiation condition for outgoing waves, $\lim_{\rhat\rightarrow \infty}\big[\rhat \pp_\rhat \phiIscn(\rhat) - \ii\xO \rhat \phiIscn(\rhat)\big] = 0$, can be written~as,
\\[-5mm]
 \bsuba{pIsc_solution}
 \bal
 \eqlab{pIscn}
 \pIscn(\rhat) & =\ii\rhoOO\omega A_n \Big[\Big(\acnsc + \Imn n(\rhat,t) \Big) \hpn n(\xO\rhat)
 \nn\\
 &\quad +\Big(\Ipn n(\infty,t)\!-\!\Ipn n(\rhat,t) \Big)\hmn n(\xO\rhat) \Big],\!
 \\
 \eqlab{I_integrals_definition}
 I_n^{(\pm)} (\rhat,t) &= \frac{a_c \xO^3}{\ii} \int_1^\rhat \xi^2 \Delta T_0(\xi,t) h_n^{(\pm)}(\xO\xi)j_n(\xO\xi) \,\dm\xi.
 \\[-8mm] \nn
 \eal
 \esuba
$\pIsc$ from \eqsref{pIsc_expansion}{pIsc_solution} reduces to the standard result when setting $\Delta T_0=0$, for which $I_n^{(\pm)} (\rhat,t)=0$ and $\acnsc = \alpha_{c,n}^\mr{sc,0}$. However, even for small temperature deviations $\DTO$ in the fluid, $\pIsc$ is significantly perturbed by thermal effects in a large region set by the thermal diffusion length, see details in the Supplemental Material~\cite{Note1}. Last, the constants $\big\{\acnsc, \asnsc, \acnp, \asnp \big\}$ are found from the boundary conditions at the particle-fluid boundary, $\rhat = 1$. The boundary conditions at the particle boundary are continuous velocity and stress,
\\[-6mm]
 \bsubal{BCs_1st_order}
 \eqlab{v1_continuous}
 \vIr &=\vIrp, &\vIth &= \vIthp,
 \\
 \eqlab{stress_continuous}
 \sigma_{1\theta r} & = \sigma_{1\theta r}' ,   &\sigma_{1rr} &= \sigma_{1rr}'.
 \esubal
In contrast to the six boundary conditions used in our work~\cite{Winckelmann2023},
we only need four here, because the adiabatic assumption excludes the thermal scalar potential
and the corresponding two thermal scattering coefficients $\atnsc,\atnp$ from the theory. Using the velocity potentials \eqsref{Helmholtz_decomposition}{standard_Helmholtz_solutions}, the stress tensors from \eqsref{sigma_fl}{sigma_sl} with the pressures from \eqssref{vIin}{pIsc_expansion}{pIsc_solution}, as well as the definition~\eqnoref{solid_definitions} of velocity and viscosity in the solid,  we write the boundary conditions~\eqnoref{BCs_1st_order} expressed for each value of $n$ as follows:
\\[-6mm]
 \bsuba{BCs_n_1st_order}
 \bal
 \eqlab{BC_n_v1r}
 &\text{\underline{$\vIr=\vIrp$}} \nn\\
 &\;\; \acnsc\xO \hpnp n(\xO)-\asnsc n(n+1)\hpn n(\xs) \nn\\
 &\;\; -\acnp\xcp j'_n(\xcp)+\asnp n(n+1) j_n(\xsp) \nn\\
 &\;\; =-\xO j'_n(\xO)-\Ipn n(\infty,t) \xO \hmnp n(\xO),
 \\[-0mm]
 \eqlab{BC_n_v1t}
 &\text{\underline{$\vIth= \vIthp$}} \nn\\
 &\;\; \acnsc \hpn n(\xO)-\asnsc \big[\xs \hpnp n(\xs)+\hpn n(\xs)\big] \nn\\
 &\;\; -\acnp j_n(\xcp)+\asnp \big[\xsp j_n'(\xsp)+j_n(\xsp)\big] \nn\\
 &\;\; = -j_n(\xO)-\Ipn n(\infty,t) \hmn n(\xO),
 \\[-0mm]
 \eqlab{BC_n_sigma1tr}
 &\text{\underline{$\sigma_{1\theta r}=\sigma_{1\theta r}'$}}
 \nn\\
 &\;\; \acnsc 2\etaO \big[\xO \hpnp n(\xO)-\hpn n(\xO)\big]
 \nn\\
 &\;\; -\asnsc \etaO \big[\xs^2 \hpnpp n(\xs)+(n^2+n-2)\hpn n(\xs)\big]
 \nn\\
 &\;\; -\acnp 2\eta_0' \big[\xcp j_n'(\xcp)\!-\!j_n(\xcp)\big]
 \nn\\
 &\;\; +\asnp \eta_0'\big[\xs^{\prime 2} j_n''(\xsp)+(n^2+n-2)j_n(\xsp)\big]
 \nn\\
 &\;\; = -2\etaO \big[\xO j_n'(\xO) -j_n(\xO)\big]
 \nn\\
 &\quad \;\; -\Ipn n(\infty,t) 2\etaO \big[\xO \hmnp n(\xO)-\hmn n(\xO)\big],
 \eal
 \bal
 \eqlab{BC_n_sigma1rr}
 &\text{\underline{$\sigma_{1rr}=\sigma_{1rr}'$}} \nn\\
 &\;\; \acnsc \eta_0 \big[(2\xO^2-\xs^2)\hpn n(\xO)+2\xO^2 \hpnpp n(\xO)\big]
 \nn\\
 &\;\; -\asnsc \eta_0 2n(n+1) \big[\xs \hpnp n (\xs)-\hpn n(\xs)\big]
 \nn\\
 &\;\; -\acnp \eta_0' \big[(2 x_\mr{c}^{\prime 2}-\xs^{\prime 2})j_n(\xcp)
 +2x_\mr{c}^{\prime 2} j_n''(\xcp) \big]
 \nn\\
 &\;\; +\asnp \eta_0' 2n(n+1) \big[ \xsp j_n'(\xsp)-j_n(\xsp) \big]
 \nn\\
 &\;\; = -\Ipn n(\infty,t) \etaO \big[(2\xO^2-\xs^2)\hmn n(\xO)+2\xO^2 \hmnpp n(\xO)\big]
 \nn\\
 &\quad \;\; -\etaO \big[(2\xO^2-\xs^2)j_n(\xO)+2\xO^2j_n''(\xO)\big] .
 \eal
 \esuba
Note that for $n=0$, \eqsref{BC_n_v1t}{BC_n_sigma1tr} are void, so in this case only two equations with two unknown coefficients need to be solved. Similar to the corresponding boundary condition equations~(52) in Ref.~\cite{Winckelmann2023}, we write \eqref{BCs_n_1st_order} as a 4-by-4 matrix equation and apply Cramer's rule to find the scattering coefficients by expanding the involved determinants to leading order in $\xO$. We use the following scalings in our derivation,
 \beq{xO_scalings}
 \Gams,I^{(\pm)}_n(\infty,t),\etaOTi^{-1} \sim \xO^2,
 \qquad
 \xcp,\xsp, \big| a_q \Delta T_0 \big| \sim \xO.
 \eeq
For solid particles, only the two scattering coefficients $\acnsc$ and $\asnsc$ of the fluid are needed to calculate $\FFFrad$. To leading order in $\xO$, the coefficient $\asnsc$ is unchanged compared to the case $\Delta T_0 =0$, whereas $\acnsc$ is,
 \beq{acnsc_form}
 \acnsc = \acnscO + \Ipn n(\infty,t),
 \eeq
where $\acnscO$ is the scattering coefficient evaluated at $\Delta T_0 =0$. The expressions for $\asnsc$ and $\acnscO$ to leading order in $\xO$ for $n=0,1,2$, which are the coefficients needed to calculate $\FFFrad$, are found to be,
 \bsubal{scattering_coefficients}
 \eqlab{ac0sc}
 \alpha_{c,0}^\mr{sc,0} &= -\frac{\ii \xO^3}{3} \big[1 - \kapSOOTi\big],
 \\
 \eqlab{ac1sc}
 \alpha_{c,1}^\mr{sc,0} &= \frac{\ii \xO^3}{3} \frac{(\rhoOOTi-1)[3(\ii\xs-1)+\xs^2]}{(2\rhoOOTi+1)\xs^2-9(1-\ii\xs)},
 \\
 \eqlab{as1sc}
 \alpha_{s,1}^\mr{sc} &= \frac{\ii\xO (\rhoOOTi-1)\xs^2\ee^{-\ii\xs}}{(2\rhoOOTi+1)\xs^2-9(1-\ii\xs)},
 \\
 \eqlab{ac2sc}
 \alpha_{c,2}^\mr{sc,0} &= \frac{2\ii\xO^5}{15} \frac{6\ii\xs^2+\xs^3-15(\ii+\xs)}{9\xs^2(\ii+\xs)},
 \\
 \eqlab{as2sc}
 \alpha_{s,2}^\mr{sc} &= -\frac{\xO^2 \xs \ee^{-\ii\xs}}{9(\ii+\xs)}.
 \esubal

\subsection{The second-order steady streaming and \textit{\textbf{F}}$^{\textbf{rad}}$}
\seclab{second_order}

We now depart from assuming an arbitrary incident axisymmetric pressure wave and focus on the important special case of a standing plane wave. The incident pressure $p_1^\mr{in}(z)$ varies spatially along the $z$-axis with wave number $\kO$, amplitude $p_a$, and phase shift $\kO d$,
 \beq{p1_in_standing}
 \pIin(z) = p_a \cos[\kO (z+d)],
 \eeq
which by comparing to \eqsref{vIin}{phiIin}, corresponds to defining the incident wave by,
 \beq{An_standing}
 A_{n} = \frac{p_a}{\ii \rhoOO \omega} \frac{2n+1}{2}\ii^{n}\Big[\ee^{\ii\kO d}+(-1)^n \ee^{-\ii\kO d}\Big].
 \eeq

To evaluate $\FFFrad$, we compute the second-order time-averaged fields, velocity $\avr{\vvvII}$ and pressure  $\avr{p_2}$. They are split into an incident (in) and a scattered (sc) part, where the former is assumed to be generated solely by the incident pressure field without any gradients in $T_0$ and with no suspended particle in the fluid,
 \bsuba{fluid_field_composition_2_and_2nd_order_incident_eqs}
 \bal
 \eqlab{fluid_pressure_2}
 \avr{p_2}&=\avr{\pIIin}+\avr{\pIIsc} ,
 \\
 \eqlab{fluid_velocity_2}
 \avr{\vvvII} &= \avr{\vvvIIin}+\avr{\vvvIIsc},
 \\
 \eqlab{cont_eq_in}
 0 &= \div\avr{\rhoOO\vvvIIin} + \frac{1}{\cOOsqr}\div\avr{\pIin\vvvIin},
 \\
 \eqlab{NS_eq_in}
 0 &= \div\Big\langle \etaOO \Big[\nablabf \vvvIIin + \big(\nablabf \vvvIIin\big)^\textsf{T}\Big]
 \nn\\
 &\quad + \big(\etaBOO - \tfrac{2}{3}\etaOO\big)\big(\div \vvvIIin\big) \,\Imat
 \nn\\
 &\quad -\pIIin \, \Imat + \etaIin \Big[\nablabf \vvvIin + (\nablabf \vvvIin)^\textsf{T}\Big]
 \nn\\
 &\quad + \big(\etaBIin - \tfrac{2}{3}\etaIin \big)(\div \vvvIin )\,\Imat -\rhoOO\vvvIin\vvvIin \Big\rangle.
 \eal
 \esuba
Subtracting \eqsref{cont_eq_in}{NS_eq_in} from \eqref{fluid_second_order}, and using that $\avr{\vvvIIin}$ and $\avr{\pIin\vvvIin}$ are negligibly small for a standing wave, we find the following equations for $\avr{\vvvIIsc}$ and $\avr{\pIIsc}$ to leading order,
 \bsubal{2nd_order_sc_eqs}
 \eqlab{cont_eq_sc}
 &\div\avr{\vvvIIsc} =  - \frac{1}{\rhoOO\cOOsqr}\div\avr{\pI\vvvI}_\mr{nii},
 \\
 \eqlab{NS_eq_sc}
 &\nuOO\nabla^2\avr{\vvvIIsc} \!+\! \big(\nuBOO \!-\! \tfrac23 \nuOO \big) \nablabf \big(\div \avr{\vvvIIsc}\big) \Imat \!-\! \frac{1}{\rhoOO}\nablabf\avr{\pIIsc} \,\Imat
 \nn\\
 &\;= -\div\Big\langle \nuI \Big[\nablabf \vvvI + (\nablabf \vvvI)^\textsf{T}\Big]
 + \Big[\nuBI - \tfrac{2}{3}\nuI \Big](\div \vvvI )\,\Imat \Big\rangle_\mr{nii}
 \nn\\
 &\;\quad + \div \avr{\vvvI\vvvI}_\mr{nii}+a_\rho \Delta T_0 \div \avr{\vvvIin\vvvIin},
 \esubal
where the index '$\mr{nii}$' (stands for 'no incident-incident') indicates that terms with products of two first-order incident fields are discarded. In \eqref{2nd_order_sc_eqs} we have introduced the kinematic viscosities to zeroth and first order,
\\[-7mm]
 \bsuba{kinematic_viscosities}
 \bal
 \eqlab{nuOnuBO}
 \nuOO &=  \frac{\etaOO}{\rhoOO} , \qquad \nuBOO = \frac{\etaBOO}{\rhoOO},
 \\
 \eqlab{nuI}
 \nuI &= \frac{\etaI}{\rhoOO} \approx - B_c^\infty \frac{\nuOO}{\rhoOO\cOOsqr}p_1,
 \\
 \eqlab{nuBI}
 \nuBI &= \frac{\etaBI}{\rhoOO} \approx - B_c^{\mr{b}\infty} \frac{\nuOO}{\rhoOO\cOOsqr}p_1.
 \eal
 \\[-6mm] \nn
 \esuba

To facilitate the computation of $\FFFrad$, we note that the incident terms, which make no reference to the particle heating or scattering, cannot contribute to the radiation force, and therefore we obtain the useful expression
\\[-5.5mm]
 \bal\eqlab{Frad_in}
 &\oint_{\partial \Omega_0} \Big\langle \etaOO \Big[\nablabf \vvvIIin + \big(\nablabf \vvvIIin\big)^\textsf{T}\Big]  + \big(\etaBOO - \tfrac{2}{3}\etaOO\big)\big(\div \vvvIIin\big) \,\Imat
 \nn\\
 & -\pIIin \, \Imat + \etaIin \Big[\nablabf \vvvIin \!+\! (\nablabf \vvvIin)^\textsf{T}\Big] + \big(\etaBIin \!- \tfrac{2}{3}\etaIin \big)(\div \vvvIin )\,\Imat
 \nn\\
 & \qquad  -\rhoOO \vvvIin\vvvIin \Big\rangle \cdot \nnn \, \dm S =0 .
 \\[-6mm] \nn
 \eal
Then, by subtracting \eqref{Frad_in} from \eqref{Frad}, we find,
\\[-7mm]
 \bal\eqlab{Frad_sc}
 &\qquad \FFFrad = \rhoOO \oint_{\partial \Omega_0} \Big\{
 \nn\\
 &
 \Big\langle \!\nuOO\! \Big[\nablabf \vvvIIsc
 \!+\! \big(\nablabf \vvvIIsc\big)^\textsf{T}\Big]
 \!+\! \big(\nuBOO \!\!-\! \tfrac{2}{3}\nuOO\big)\big(\div \vvvIIsc\big)\Imat
 \!-\! \tfrac{1}{\rhoOO}\pIIsc \Imat \Big\rangle
 \nn\\
 & +  \Big\langle \nuI\! \Big[\nablabf \vvvI \!+\! (\nablabf \vvvI)^\textsf{T}\Big] + \big(\nuBI - \tfrac{2}{3}\nuI \big)(\div \vvvI )\,\Imat - \vvvI\vvvI \Big\rangle_\mr{nii}
 \nn\\
 & \qquad  -a_\rho \Delta T_0 \avr{\vvvIin\vvvIin} \Big\} \cdot \nnn \, \dm S =0.
 \\[-8mm] \nn
 \eal
The set of equations~\eqsnoref{2nd_order_sc_eqs}{Frad_sc} determines $\FFFrad$ similarly to Eqs.~(17) and (82) in Ref.~\cite{Winckelmann2023} (setting $\avr{\vvvIIin}=0$), but differing by the appearance here of the additional terms containing $\avr{\vvvIin\vvvIin}$. However, the solution method remains unchanged, as these new terms are included using the same Helmholtz decomposition and the same partial-wave expansion as for the other terms containing products of first-order fields. From this point, $\avr{\vvvIIsc}$, $\avr{\pIIsc}$, and $\FFFrad$ are thus computed using the same technique as detailed in Ref.~\cite{Winckelmann2023}, Sections IV~A, IV~B, V~A, and Appendix D, and therefore it suffices here simply to summarize the result for $\FFFrad$. Since $\avr{\vvvIIin}$ is neglected here, $\FFFrad$ only contains the component consisting of time-averaged first-order products, denoted $\FFFradII$ in Ref.~\cite{Winckelmann2023} Eqs.~(84) and~(87), and not the drag-force component, denoted $\FFFradin$~\cite{Winckelmann2023},
\\[-15mm]
\begin{widetext}
\vspace*{-6mm}

 \bal\eqlab{Frad_full}
 \FFFrad &=  -\eee_z 3\pi\rhoOO \bigg\{
 -\frac{a^2 \nuOO}{\rhoOO \cOOsqr} \int_1^\infty \dm\xi
 \int_0^\pi \dm\theta\: \cos\theta\sin\theta \;
 \div \frac{1}{2}\re\big[p_1 \vvvI \big]_\mr{nii}
 \\
 &+ a^2 \int_1^\infty \dm\xi
 \int_0^\pi \dm\theta \: \sin\theta
 \big(1-\xi^{-2}\big) \frac{1}{2}\re\bigg[\eee_r\cdot \Big(a_\rho \Delta T_0
 \vvvIin \vvvI^{\mr{in}*} \Big)  \cdot \eee_z + \frac{1}{2}a\xi \Big\{a_\rho \Delta T_0 \div \Big(\vvvIin \vvvI^{\mr{in}*} \Big)\Big\} \cdot \eee_\theta \sin\theta
 \nn
 \\
 &\hspace*{20mm} +\eee_r\cdot \Big(
 \vvv_1 \vvv_1^* -\nuI\Big[\nablabf \vvv_1+(\nablabf \vvv_1)^\textsf{T}\Big]^*
 -\Big[\nuIB-\tfrac{2}{3}\nuI\Big](\div \vvv_1^*) \Imat \Big)_\mr{nii}  \cdot \eee_z
 \nn
 \\
 &\hspace*{20mm} +\frac{1}{2}a\xi \Big\{ \div \Big( \vvv_1 \vvv_1^*
 -\nuI\Big[\nablabf \vvv_1+(\nablabf \vvv_1)^\textsf{T}\Big]^*
 -\Big[\nuIB-\tfrac{2}{3}\nuI\Big](\div \vvv_1^*) \Imat \Big)_\mr{nii}\Big\} \cdot \eee_\theta \sin\theta
 \bigg]
 \nn\\
 \nn
  &+\int_0^\pi  \!\dm\theta\:\frac{\sin\theta}{2}\re\bigg[\frac{a^3}{\xs^2}\Big(\vIr\pp_r \vIr^*+\frac{1}{r}\vIth \pp_\theta \vIr^* -\frac{1}{r}\vIth\vIth^*\Big)\cos\theta
 -\frac{a^3}{\xs^2}\Big(\vIr\pp_r \vIth^*+\frac{1}{r}\vIth\pp_\theta \vIth^* -\frac{1}{r}\vIr\vIth^* \Big)\sin\theta\bigg]_{\rhat=1}
 \bigg\}.
 \eal
\end{widetext}
We emphasize two differences between this expression for $\FFFrad$ and expression~(87) in Ref.~\cite{Winckelmann2023} for $\FFFradII$. First, two extra terms containing $a_\rho \Delta T_0\vvvIin \vvvI^{\mr{in}*}$ enter in the second integral of \eqref{Frad_full}. Second, the first-order fields here are different due to the inclusion of a spatially varying $T_0$ through the integrals $I_n^{(\pm)}(\rhat,t)$ and their $\rhat$-derivatives, and due to the assumption of an adiabatic acoustic wave which was not enforced in Ref.~\cite{Winckelmann2023}. Because of the latter, the final result here does not contain any of the combinations of thermal scattering coefficients $\atn$ that appear in $\FFFradII$ of Ref.~\cite{Winckelmann2023}. The mathematical structure of $\FFFrad$ is a sum over quadratic terms of scattering coefficients,
 \bsubal{F1_form_sl}
 \eqlab{F1_expr_sl}
 \FFFrad &= -\eee_z 3\pi \rhoOO \sum_{n=0}^\infty \frac{n+1}{(2n+1)(2n+3)}\re(A_nA_{n+1}^*D_n),
 \nn \\
 &\text{ with }\; D_n = D_n^0 + D_n^{\Delta T_0},
 \\
 \eqlab{Dn_expr}
 D_n^i &= S_{00,n}^i + S_{0c,n}^i \alpha^\mr{sc,0 *}_{c,n+1} +  S_{0s,n}^i \alpha^\mr{sc *}_{s,n+1} + S_{c0,n}^i \alpha^\mr{sc,0}_{c,n}
 \nn\\
 & \quad + S_{cc,n}^i \alpha^\mr{sc,0}_{c,n} \alpha^\mr{sc,0 *}_{c,n+1}
 + S_{cs,n}^i \alpha^\mr{sc,0}_{c,n} \alpha^\mr{sc *}_{s,n+1}
 + S_{s0,n}^i \alpha^\mr{sc}_{s,n}
 \nn\\
 &\quad + S_{sc,n}^i \alpha^\mr{sc}_{s,n} \alpha^\mr{sc,0 *}_{c,n+1}
 + S_{ss,n}^i \alpha^\mr{sc}_{s,n} \alpha^\mr{sc *}_{s,n+1}, \!\! \quad i=0,\Delta T_0.
 \esubal
Here, we have defined the force coefficients $D_n$ and the second-order coefficients $S_{kl,n}^i$, $i=0,\Delta T_0$. The force coefficients have been split into the two components $D_n^0$ and $D_n^{\Delta T_0}$, where $D^0_n$ are the force coefficients for $\Delta T_0 = 0$, and $D_n^{\Delta T_0}$ contains terms that depend on $\Delta T_0$ explicitly or implicitly  through $I_n^{(\pm)}(\rhat,t)$. To determine the the second-order coefficients $S_{kl,n}^i$, we insert $p_1$ from \eqssref{vIin}{pIsc_expansion}{pIsc_solution} and $\vvvI$ from \eqsref{vIin}{vIsc} with the potentials from \eqref{standard_Helmholtz_solutions} and read off the coefficients in front of each combination of the scattering coefficients. To leading order in the small parameter $\xO$, only $D_0^0$, $D_1^0$, and $D_n^{\Delta T_0}$ contribute to $\FFFrad$. Further, the only leading-order contribution to $D_n^{\Delta T_0}$ arise from terms in $S_{00,n}^{\Delta T_0}$ that scale with $\xO^2a_c\Delta T_0$. To leading order we obtain
 \bal
 \eqlab{DnDeltaT0}
 &D_n^{\Delta T_0}(t) = \!\int_1^\infty \!\!\Big(1-\frac{1}{\rhat^{2}}\Big)
 \bigg\{\! -2 a_c \Delta T_0 \xO^2 j_n(\xO \rhat) j_{n+1}(\xO \rhat)
 \nn\\
 & +\frac{\ii}{\rhat^2}\Big[\Ipn n(\rhat,t) \!+\! \Imn n(\rhat,t) \!-\! \Ipn{n+1}(\rhat,t) \!-\! \Imn{n+1}(\rhat,t)\Big] \bigg\} \; \dm \rhat
 \nn\\
 & = 2a_c\xO^2\int_1^\infty \Delta T_0 \bigg[\xO\bigg(\rhat-\frac{1}{3\rhat}\bigg) \Big(j_n^2(\xO \rhat) - j_{n+1}^2(\xO \rhat)\Big)
 \nn\\
 & \hspace{14mm} -\Big(1-\frac{1}{\rhat^{2}}\Big) j_n(\xO \rhat) j_{n+1}(\xO \rhat) \bigg] \; \dm \rhat,
 \eal
where \eqref{I_integrals_definition} and integration by parts was used to reach the final expression.
All the coefficients $D_n^{\Delta T_0}$ scale with $\xO^2a_c\Delta T_0$, although they decrease in magnitude with $n$, and one may need to evaluate many terms in the sum of \eqref{F1_form_sl} to reach convergence. We note that $D_n^{\Delta T_0}$ does not depend on the boundary-layer thickness $\dels$, and thus it can cause significant contributions to $\FFFrad$, without entering the viscous limit of $\dels \gtrsim a$. In \appref{S_coeff} we list all the coefficients $S_{ik,0}^0$ and $S_{ik,1}^0$ needed to compute $D_0^0$ and $D_1^0$. The final expression in the long wavelength limit for $\FFFrad$ on a heated particle in an incident standing pressure wave is,
 \bal
 \eqlab{FradLongwave}
 \FFFrad &= -\eee_z\:\pi \rhoOO \re\bigg(A_0A_1^*D_0^0+\frac{2}{5}A_1A_2^* D_1^0\bigg)
 \\
 &-\eee_z 3\pi \rhoOO \sum_{n=0}^\infty \frac{n+1}{(2n+1)(2n+3)}\re\big(A_nA_{n+1}^*D_n^{\Delta T_0}\big),
 \nn \\
 & \text{\makebox[2.5em][l]{with}}
 D_0^0 \text{ and } D_1^0 \text{ from \eqref{Dn_expr}},
 \nn\\
 & \text{\makebox[2.5em][l]{}} \alpha^\mr{sc,0}_{c,n} \text{ for } n=0,1,2 \text{ from \eqref{scattering_coefficients}},
 \nn\\
 & \text{\makebox[2.5em][l]{}} \alpha^\mr{sc}_{s,n} \text{ for } n=1,2 \text{ from \eqref{scattering_coefficients}},
 \nn\\
 & \text{\makebox[2.5em][l]{}} S_{ik,n}^0 \text{ for } n=0,1 \text{ from \appref{S_coeff}},
 \nn\\
 & \text{\makebox[2.5em][l]{}} D_n^{\Delta T_0}  \text{ from \eqref{DnDeltaT0}, and }
   A_n \text{ from \eqref{An_standing}}.
 \nn
 \eal
Finally, using \eqref{T0_ana} we approximate $D_n^{\Delta T_0}$ in the limit $t\rightarrow \infty$ and find the following expressions for $\Delta T_0$,
 \beq{delT0_limit}
 \DTO(\rhat,\infty) = \frac{1}{\rhat}\:\DTOsurf.
 \eeq
This result combined with \eqref{DnDeltaT0}  leads to the following expression in leading order,
 \bal\eqlab{DnDeltaT0_approx}
 &D_n^{\DTO}(\infty) =  \frac{\pi}{(2n+1)(2n+3)}\xO^2 a_c \DTOsurf .
 \eal

\section{Results for a standing plane wave}
\seclab{ResultsPlaneWave}

Inserting $A_n$ from \eqref{An_standing} in \eqref{FradLongwave}, we obtain $\FFFrad$ from a standing plane wave on a spherical particle of radius $a$ to leading order $\xO^3 = \kO^3a^3$ or $\xO^2a_c\:\DTO$,
 \bsuba{Frad_standing}
 \bal
 \eqlab{Frad_standing_form}
 \FFFrad &= 4\pi \Phiac a^3 k_0 \Eac \sin(2k_0d)\:\een_z,
 \\
 \eqlab{Eac}
 \Eac & = \tfrac{1}{4}\kapSOO p_a^2,
 \\
 \eqlab{Phiac}
 \Phiac(t) &= \Phiac^0+\Phiac^{\DTO}(t) ,
 \\
 \eqlab{PhiacO}
 \Phiac^0 &= \tfrac{3}{2}\:x_0^{-3}\re\!\Big(D_0^0-2D_1^0\Big),
 \\
 \eqlab{PhiacDelT0}
 \Phiac^{\DTO}(t) &= \tfrac{3}{2}\:x_0^{-3}\sum_{n=0}^{\infty}(n+1)
 \re\!\Big[(-1)^n D_n^{\DTO}(t)\Big],
 \eal
 \esuba
where we have introduced the usual time-averaged acoustic energy density $\Eac$ and acoustic contrast factor $\Phiac$ \cite{King1934, Yosioka1955, Settnes2012, Karlsen2015}. The latter is split as $\Phiac = \Phiac^0 + \Phiac^{\DTO}(t)$ into the sum of an ambient-temperature term $\Phiac^0$ with $\DTO =0$ and a term $\Phiac^{\DTO}(t)$ due to the particle heating $\DTO >0$. A main feature is that heating through $\DTO$ may cause a sign reversal of $\Phiac$, thus possibly reversing the direction of particle focusing from pressure nodes to anti-node, or \textit{vice versa}. We obtain the asymptotic limit $\Phiac^{\DTO}(\infty)$ for long times of the acoustic contrast-factor perturbation $\Phiac^{\DTO}(t)$ by combining \eqsref{DnDeltaT0_approx}{PhiacDelT0},

 \beq{PhiacDelT0inf}
 \Phiac^{\DTO}(\infty)=\frac{3\pi}{8} \frac{a_c \DTOsurf}{\xO}.
 \eeq
The time evolution of the acoustic contrast factor $\Phiac(t)$ is found numerically by evaluating $D_n^{\DTO}(t)$ in \eqref{DnDeltaT0}, with $\DTO(t)$ from \eqref{T0_ana} for a polystyrene particle of radius $a=1$, 2, and $5~\SImum$ in the three different liquids (a) water, (b) ethanol, and (c) oil. We have  chosen the heat power density $P_0 = 3 \DTOsurf \kthOO a^{-2}$ such that the asymptotic surface temperature increase becomes $\DTOsurf = 1~\SIK$. The values of the power density $P_0$ and the absorbed power $\dot{Q}=\tfrac{4\pi}{3}a^3 P_0$ are listed in \tabref{PO_table}, and the material parameters are given in \tabref{Param_table}. The resulting $\Phiac$ as a function of time is shown in \figref{Phiac}, where we have used the terms $n \leq 15$ in the sum of \eqref{PhiacDelT0} to reach a satisfactory convergence.

\begin{table}[t]
\caption{\tablab{PO_table} Power density $P_0$ and absorbed power $\dot{Q}$ for the nine cases shown in \figref{Phiac}: a polystyrene particle of radius $a = 1$, 2, and $5~\SImum$ in water, oil, and ethanol, respectively.}
\begin{ruledtabular}
\begin{tabular}{cccccc}
$a$  & Source &  Water  &  Ethanol  &  Oil & Unit
\\ \hline
$1\,\SImum$  & $P_0$ & $1800$  & $500$  & $500$ & $\SIGW\:\SIm^{-3}$ \upspace
\\
$2\,\SImum$  & $P_0$ & $460$  & $130$  & $120$ & $\SIGW\:\SIm^{-3}$
\\
$5\,\SImum$  & $P_0$ & $73$  & $20$  & $20$ & $\SIGW\:\SIm^{-3}$
\\ \hline
$1\,\SImum$  & $\dot{Q}$ & $7.7$  & $2.1$  & $2.1$ & $\SImuW$ \upspace
\\
$2\,\SImum$  & $\dot{Q}$ & $15$  & $4.1$  & $4.2$ & $\SImuW$
\\
$5\,\SImum$  & $\dot{Q}$ & $38$  & $11$  & $10$ & $\SImuW$
\\
\end{tabular}
\end{ruledtabular}
\end{table}

\begin{table}[t]
\caption{\tablab{Param_table} Parameters at $T_0^\infty=300$~K for water \cite{Muller2014, Wagner2002, Huber2009, Huber2012},  oil \cite{Coupland1997, Noureddini1992, Ghosh2017}, ethanol \cite{Sun1988, Dukhin2009, Cheric_ethanol}, and polystyrene \cite{OndaCorp, CRCpolymers, Domalski1996, Chang1968, Smith1972, Karlsen2015} used in the examples in \figref{Phiac}. We list the parameters necessary to compute $\DTO$ from \eqref{T0_ana}, the scattering coefficients $\ain$ from \eqref{scattering_coefficients}, and the second-order coefficients $S_{ik,n}$ found in appendix \appref{S_coeff}. Note that $\kapSO$ can be found from \eqsref{compressibility_fl}{compressibility_sl} for a fluid and a solid, respectively. Due to lack of data, we set $\big(\frac{\pp\eta}{\pp \rho}\big)_{\rho_0} = 0$ for oil and ethanol.}
\begin{ruledtabular}
\begin{tabular}{lcccccl}
Para- & Water  & Oil  & Etha- & Poly-    & Unit \\
meter &        &      & nol   & styrene  &     \\
\hline
$\cO,\cLO$      & $1502$  & $1445$   & 1138 &    2407   & $\mr{m\,s^{-1}}$
\\
$\cTO$ & -- & -- & -- & 1154 & $\mr{m\,s^{-1}}$
\\
$\rhoO$  & $996.6$  & $922.6$   & 784 &  1050 & $\SIkgm$
\\
$\alphapO$   & $0.275$   & $0.705$ &  $1.104$ & $0.209$ & $10^{-3}\: \SIK^{-1}$
\\
$\cpO$  & $4181$   & $2058.4$  & 2445 &  1241  & $\SIJ\:(\SIkg\:\SIK)^{-1}$
\\
$\kthO$   & $0.61$   & $0.166$  & 0.167 & 0.154  & $\SIW\:(\SIK\:\SIm)^{-1}$
\\
$\DthO$   & $1.464$   & $0.874$  & $0.871$ & $1.182$  & $10^{-7}\: \SIm^{2}\:\SIs^{-1}$
\\
$\gamO$
& 1.012 & 1.15 & 1.19 & 1.04 & --
\\
$\etaO$  & $0.854$   & $57.4$  & $1.01$ & --  & $\SImPas$
\\
$\etaO^\mr{b}$  & $2.4$ & $85.13$  & $1.4$ & -- & $\SImPas$
\\
$\frac{1}{\etaO}\frac{\pp \eta}{\pp T}\big|_0$   & $-0.022$  & $-0.044$ & $-0.019$ & --  & $\SIK^{-1}$
\\
$\frac{1}{\etaO}\frac{\pp\eta}{\pp \rho}\big|_0$ & $-2.3\times 10^{-4}$ & --  & -- & --  & $\SIm^3\:\SIkg^{-1}$
\\
$a_c$, \eqnoref{ac_arho}  & $1.7$ & $-2.2$  & $-3.0$ & --  &  $10^{-3}\: \SIK^{-1}$
\\
\end{tabular}
\end{ruledtabular}
\end{table}

For a polystyrene particle  in water, we see in \figref{Phiac}(a) that for all three particle sizes the initial value of the acoustic contrast factor is nearly the same, $\Phiac(0) = 0.17$. Even for the largest particle, $a = 5~\SImum$, the heating induces a 59\% increase, $\Phiac(\infty) = 0.27$, and this effect increases to $\Phiac(\infty) = 0.41$ (141\% increase) and $0.64$ (276\% increase) as the particle radius decreases to $a = 2$ and $1~\SImum$, respectively.

\begin{figure}[t]
\centering
\includegraphics[width=0.95\linewidth]{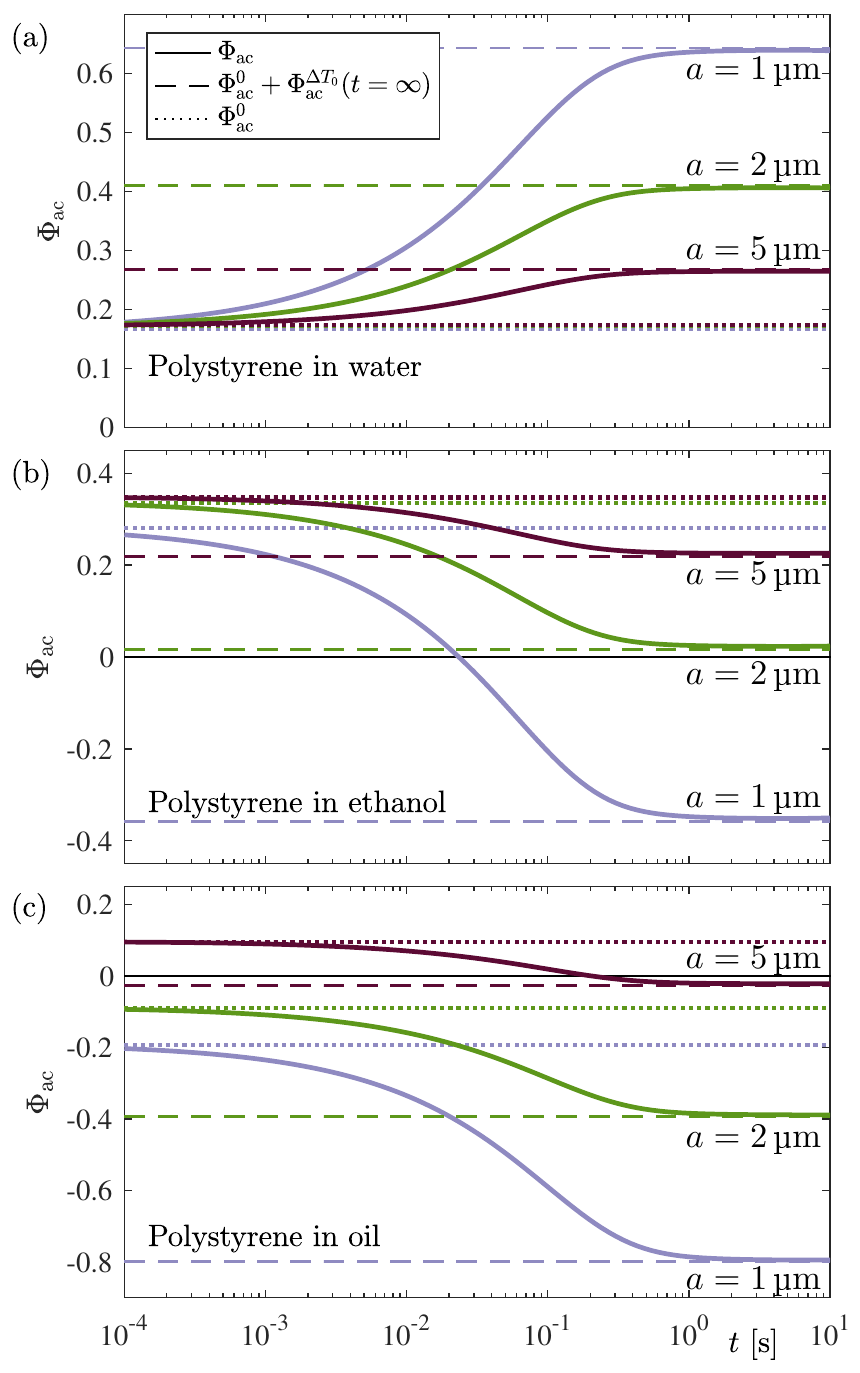}
\caption{\figlab{Phiac} The acoustic contrast factor $\Phiac$ at frequency $f=1~\SIMHz$ plotted as function of time $t$ for a polystyrene particle with radius $a=1$, $2$, or $5\,\SImum$ heated to $\DTOsurf=1~\SIK$ in different liquids: (a) water with $t^\mr{diff}_\lambda = 2.6~\SIs$, (b) ethanol with $t^\mr{diff}_\lambda = 2.5~\SIs$, and (c) oil with $t^\mr{diff}_\lambda = 4.0~\SIs$, where $t^\mr{diff}_\lambda =\lambda^2/(6 \DthO)$ is the heat diffusion time across one wavelength $\lambda$.}
\end{figure}

For the two organic liquids ethanol and oil, we see a much stronger effect in \figref{Phiac}(b,c). First, we notice that $\Phiac(0)$ depends on radius before the onset of the heating, a well-known effect due to the viscous boundary layer, in line with the previous studies of Refs.~\cite{Doinikov1994, Doinikov1994a, Doinikov1997, Doinikov1997a, Doinikov1997b, Settnes2012, Karlsen2015, Winckelmann2023}:
As radius decreases, $a = 5$, 2, and $1~\SImum$, we find the decreasing values $\Phiac(0) = 0.35$, 0.34, and 0.28 for ethanol, and more pronounced, and even with a sign change, the values $\Phiac(0) = 0.09$, $-0.09$, and $-0.19$ for oil. Second, we find that the heating of the particle induces even larger changes in the acoustic contrast factor. For ethanol, heating leads to $\Phiac(\infty) =  0.22$, $0.02$, and $-0.36$ for decreasing radius  $a = 5$, 2, and $1~\SImum$, and correspondingly for oil,  $\Phiac(\infty) =  -0.03$, $-0.39$, and $-0.80$.

The observed heat-induced increase $\Phiac^{\DTO}(t)$ in $\Phiac$ for water and decrease for ethanol and oil is due to the opposite signs of the thermal sound-speed coefficient $a_c$, \eqref{ac_arho}, of the respective sound speeds $\cO$. The effect is so large that we predict a sign reversal in $\FFFrad$ for a polystyrene particle with $a = 1~\SImum$ in ethanol and $a = 5~\SImum$ in oil. Further, $\Phiac^{\DTO}(t)$ is established by bulk dynamics on a timescale around 1~s, after which it is well approximated by \eqref{PhiacDelT0inf}. The physical mechanism causing this change in $\FFFrad$ is the ensuing heating of the bulk fluid by the heated particle that changes the scattered waves defined in Eqs.~\eqnoref{fluid_field_decomposition} - \eqnoref{Scattered_equations}, which in turn induce the acoustic microstreaming that generates a drag force on the particle, as sketched in \figref{problem_sketch}.

\begin{figure}[t]
\centering
\includegraphics[width=0.95\linewidth]{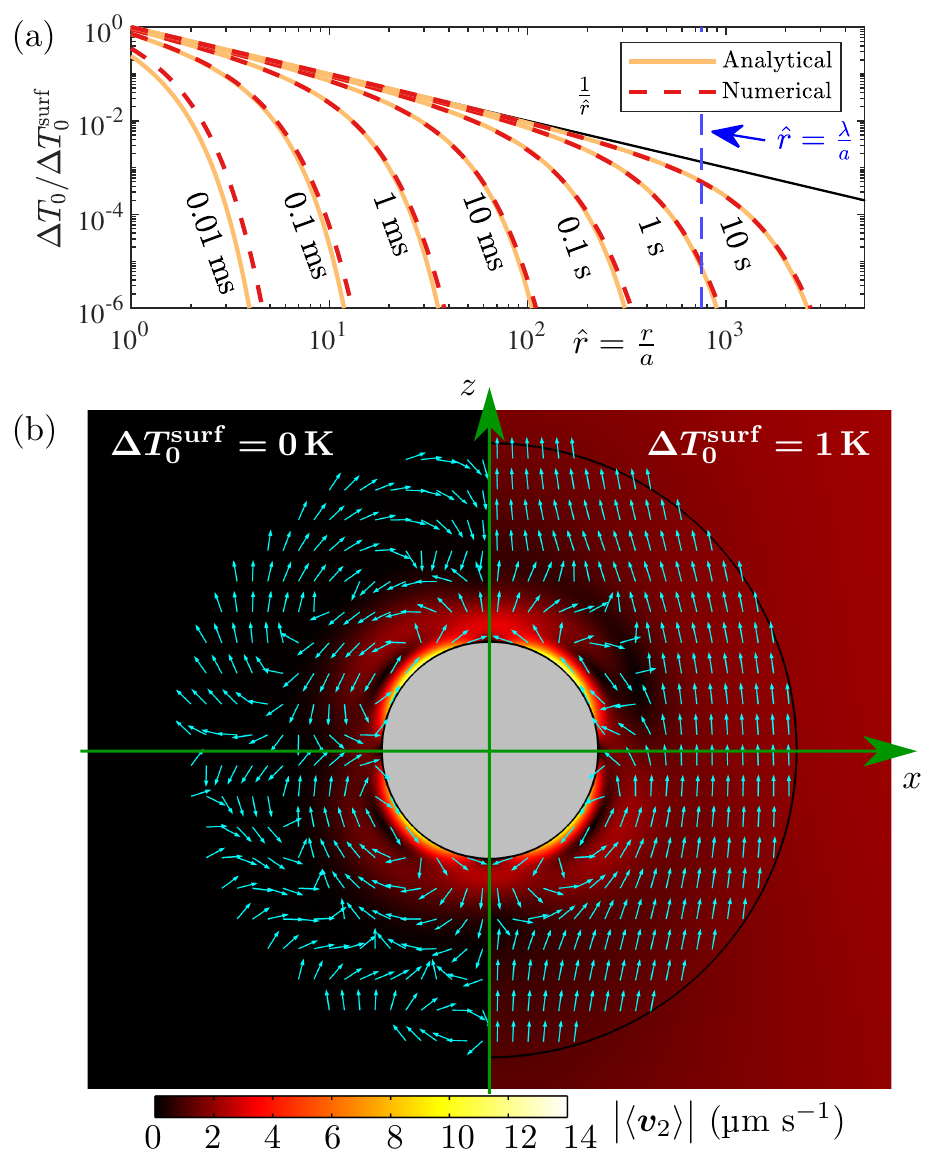}
\caption{\figlab{DelT0_v2} Results for a polystyrene particle of radius $a=2~\SImum$ in water in a standing plane wave at 1~MHz (wavelength $\lambda = 1.5~\SImm$) and at $\TOO = 300~\SIK$. (a) The time evolution of the radial temperature deviation $\DTO(\hat{r},t)$ in the fluid for $1<\hat{r} < 5000$ computed analytically (orange line) from \eqref{delT0_limit} and numerically (red dashed line).  (b) A unit-vector plot of the direction of the streaming velocity $\avr{\vvvII}$ and a color plot of its amplitude $|\avr{\vvvII}|$ for $\DTO^\mr{surf}=0~\SIK$ (left half) and $\DTO^\mr{surf}=1~\SIK$ (right half) computed numerically in \textsc{Comsol Multiphysics}.}
\end{figure}

More details of the physical mechanism causing the heat-induced $\FFFrad$ are illustrated in \figref{DelT0_v2} for the case of a 2-$\SImum$-radius polystyrene particle suspended in water, subject to a 1-MHz standing, plane, ultrasound pressure wave \eqref{p1_in_standing} of amplitude $p_a = 0.1~\SIMPa$ and phase shift $\kO d = \frac13\pi$, and heated to a surface temperature $\DTOsurf = 1~\SIK$ above the ambient temperature of 300~K. The numerical results were obtained in \textsc{Comsol Multiphysics} \cite{Comsol60} as described in the Supplemental Material~\cite{Note1}.

The time development of the temperature deviation $\DTO(\hat{r},t)$ \eqref{T0_assumptions} in the fluid from ambient temperature, caused by heating of the particle, is plotted in \figref{DelT0_v2}(a). The numerical results show that the analytical long-time-limit expression~\eqnoref{delT0_limit} is a good approximation already after $\sim1~\SIms$. This result constitutes a numerical validation of the model, and it establishes the range of validity of the analytical expression \eqref{delT0_limit} for $\DTO(\hat{r},t)$. Moreover, it is seen that after $\sim10$~s, $\DTO$ has developed into the stationary $\rhat^{-1}$-form in the one-wavelength region $r < \lambda$. This timescale is the same as the heat-diffusion timescale $t^\mr{diff}_\lambda = \lambda^2/(6 \Dth)$ observed in \figref{Phiac} for $\Phiac$ to reach its asymptotic value $\Phiac^{\DTO}(\infty)$, \eqref{PhiacDelT0inf}.

The heat-induced change to the acoustic microstreaming $\avr{\vvvII}$ around the particle is illustrated in \figref{DelT0_v2}(b). In general at zero heating, $\avr{\vvvII}$ contains several multipole components, as shown by the unit-vector plot on the left half of the figure ($\DTOsurf = 0~\SIK$). Remarkably, the heating of the particle results in a strong enhancement of a unidirectional (dipole) component as shown in the right half of the figure ($\DTOsurf = 1~\SIK$). It is this change in morphology of the microstreaming field  $\avr{\vvvII}$ that causes the heating-induced change of the force coefficient $D^{\DTO}_n$ in $\FFFrad$, \eqref{FradLongwave}, and the corresponding change $\Phiac^{\DTO}$ of the contrast factor,  \eqref{PhiacDelT0}. The amplitude and direction of the unidirectional component in the microstreaming correlates with the amplitude and sign of $\Phiac^{\DTO}$, exemplified by the nine examples plotted in \figref{Phiac}.

A key assumption of our analysis is that the heated particle does not execute a significant time-averaged motion, such that the temperature field $\DTO(\rrr,t)$ can develop in the surrounding fluid from a stationary heat source. Since the heat-induced perturbation in $\FFFrad$ is established on a timescale around $1~\SIs$, see \figref{Phiac}, our analysis is primarily relevant for determining the final equilibrium position of the heated particle. The dynamics of slowly-moving particles may be described by the theory, however the description becomes increasingly inaccurate in the transient phase the faster the particle is moving. Regardless of such inaccuracies in the detailed dynamics, our model has a robust prediction regarding whether the particle migrates to a node ($\Phiac > 0$) or an antinode ($\Phiac < 0$) in the pressure field as heating of the particle may change the sign of $\Phiac$ as seen for the 1-$\SImum$-radius polystyrene particle in ethanol and the 5-$\SImum$-radius polystyrene particle in  oil in \figref{Phiac}(b) and (c), respectively. The detailed dynamics only affects the time it takes the particle to reach its temperature-dependent equilibrium position. By combining particle heating with tuning of the solute by introducing various solvents, other cases of sign reversal in acoustophoresis may be obtained. One example is to add iodixanol to an aqueous solution of particles with an initial positive contrast factor to obtain a negative contrast factor, the so-called medium-tuning technique~\cite{Augustsson2016}. Subsequently, a second sign reversal may be obtained by heating the particle.

\section{Conclusion}
\seclab{Conclusion}

We have derived an analytical theory for the acoustic radiation force $\FFFrad$ on a heated spherical solid particle in an incident standing plane wave. The theory assumes that the external or internal heating of the particle is low enough that only small perturbations in the physical parameters of the solid particle and the external fluid occur. Further, effects of thermal convection are assumed negligible. In \secref{zeroth_order} we analyze how the temperature increase $\DTO$ diffuses from the heated particle into the surrounding fluid, and we derive the analytical expression~\eqnoref{T0_ana}; in \secref{first_order} we compute how an incident pressure wave scatters on the slightly heated sphere; and in \secref{second_order} we derive the main result of our work, the general expression~\eqnoref{F1_expr_sl} for $\FFFrad$ in terms of the force coefficients $D_n^i$ \eqref{Dn_expr}, as well as the analytical expression~\eqnoref{FradLongwave} for $\FFFrad$ in the long-wavelength limit. Furthermore, in relation to \eqref{DnDeltaT0}, we point out that the heat-induced change $D_n^{\Delta T_0}$ in the force coefficient $D_n$ does not depend on the boundary-layer thickness $\dels$, but instead on the scattering induced by the long-range, $1/r$-decaying, thermal changes in the bulk acoustic properties of the fluid. Consequently, $\FFFrad$ can be greatly perturbed by particle heating even for particles of large radius $a \gtrsim \dels$ in the long-wavelength limit, $a \ll \lambda$.

In \secref{ResultsPlaneWave}, we analyze the changes to $\FFFrad$ in an incident standing plane wave, by studying numerically and analytically the changes $\Phiac^{\DTO}$ to the acoustic contrast factor due to the temperature increase $\DTO$ caused by the heated particle. We find in \figref{Phiac} that by heating a polystyrene particle of radius $a = 1$, 2, and $5~\SImum$ suspended in water, ethanol, or oil, significant quantitative (up to an order of magnitude) and even qualitative changes (sign reversal) to the acoustic contrast factor $\Phiac$ occur. We point out that the opposite sign in the observed heat-induced change $\Phiac^{\DTO}(t)$ of $\Phiac$ for water and for the two organic liquids ethanol and oil is due to the opposite signs of the thermal sound-speed coefficient $a_c$, see \eqref{ac_arho} and \tabref{Param_table}.

The crucial role of microstreaming in causing the observed heat-induced changes in $\FFFrad$ is illustrated in \figref{DelT0_v2}(b). The analysis shows how a unidirectional component in the microstreaming is strongly enhanced by the heating of the fluid surrounding the heated particle. The drag force resulting from this microstreaming component is a main cause of the resulting changes of the acoustic radiation force $\FFFrad$.

We have extended the analytical theory of the acoustic radiation force on single spherical solid particle suspended in a homogeneous Newtonian fluid, to take into account heating of the particle. Examples where particle heating may greatly affect the force on the particle are explored. The heat-induced change in the acoustic contrast factor found in this work provides an additional control parameter for acoustofluidic handling of suspended microparticles. We speculate that this control may be obtained by optical methods such as absorbtion of laser-light by dyed particles. We hope that the presented analysis will inspire experimental efforts in the field of microscale acoustofluidics trying to improve particle-sorting, -separation, and -trapping techniques based on our predictions.

\appendix
\begin{widetext}

\section{Approximate temperature profile around uniformly heated sphere}
\seclab{T0_approximation}
The solution to the heat diffusion problem described in \secref{zeroth_order} can be found in Ref.~\cite{Goldenberg1952}, and adapted to the notation used in this work one has,
 \bsuba{T0_solution_full}
 \bal
 \eqlab{T0p_full}
 \DTO'(\rhat,t) &= \frac{Pa^2}{3k_0^{\mr{th}\prime\infty}}\Bigg[\kthOOTi + \frac12 \Big(1-\rhat^2\Big) - \frac{6}{\pi \rhat}\sqrt{\frac{1}{\rhoOOTi \cpOOTi \kthOOTi}} \int_0^\infty \frac{\ee^{-\xi^2 t/\td}}{\xi^2} \frac{\big[\sin\xi-\xi\cos\xi\big]\sin\!\big(\xi\rhat\big)}{\big[g(\xi)\big]^2 + \frac{1}{\rhoOOTi \cpOOTi \kthOOTi} \xi^2 \sin^2\xi } \,\dm\xi \Bigg] ,
 \\
 \eqlab{T0_full}
 \DTO(\rhat,t) &= \frac{P_0 a^2}{3\kthOO \rhat} \Bigg[1- \frac{6}{\pi}\int_0^\infty \frac{\ee^{-\xi^2 t/\td}}{\xi^3} K(\xi,\rhat) \,\dm\xi\Bigg] ,
 \\
 \eqlab{K_func}
 K(\xi,\rhat) &= \frac{\big[\sin\xi-\xi\cos\xi\big]}{\kthOOTi} \frac{ \sqrt{\frac{1}{\rhoOOTi \cpOOTi \kthOOTi}} \xi\sin\xi \cos\!\bigg[\sqrt{\tilde{D}_0^{\mr{th}\infty}}\xi\big(\rhat-1\big)\bigg]
 -g(\xi) \sin\!\bigg[\sqrt{\tilde{D}_0^{\mr{th}\infty}}\xi\big(\rhat-1\big)\bigg] }
 {\big[g(\xi)\big]^2 + \frac{1}{\rhoOOTi \cpOOTi \kthOOTi} \xi^2 \sin^2\xi } ,
 \\
 \eqlab{g_function}
 g(\xi) &= \Big[1-\big(\kthOOTi\big)^{-1}\Big]\sin\xi-\xi\cos\xi  .
 \eal
 \esuba
For $t\gtrsim 5 \td$, the function $\frac{\ee^{-\xi^2 t/\td}}{\xi^3}$ decays so rapidly that the integral in \eqref{T0_full} is well approximated by using the leading order in $\xi$ of $K(\xi,\rhat)$, $K(\xi,\rhat) \approx \frac{\xi^2}{3}\sin\!\Big(\sqrt{\tilde{D}_0^{\mr{th}\infty}}\xi\rhat\Big)$, where we have kept in mind that $\xi\rhat$ can still be large. Using
 \beq{error_integral}
 \frac{6}{\pi}\int_0^\infty \frac{\ee^{-\xi^2 t/\td}}{3\xi} \sin\!\Big(\sqrt{\tilde{D}_0^{\mr{th}\infty}}\xi\rhat\Big) \,\dm\xi = \mr{erf}\big(\xth \rhat\big),
 \eeq
we arrive at the result in \eqref{T0_ana}.

\section{The second-order coefficients \textit{\textbf{S}}$_{\textit{\!\textbf{ik,n}}}$ for a solid particle in a fluid}
\seclab{S_coeff}

\vspace*{-1mm}

The 13 $\Smn{ik}n$ coefficients that contribute to $D_n^0$ to leading order in $\xO$ are stated: 7 coefficients for $n=0$ and $6$ for $n=1$. For the remaining $16$ coefficients in modes $n=0$ and $n=1$, we only state their order in $\xO$ here.

 \bal\eqlab{S_coeff_sl}
 \Smn{00}0 &= \frac{x_0^3}{3\xs^2}, \qquad
 \Smn{00}1 = \frac{x_0^3}{3\xs^2}, \qquad
 \Smn{0c}0 = \frac{2\ii}{3},  \qquad
 \Smn{0c}1 \sim \calO (1) , \qquad
 \Smn{0s}0 = -x_0^2\frac{2\ii (1+B_c^\infty)}{\xs^2}\ee^{-\xs},\qquad
 \Smn{0s}1 \sim \calO(x_0^3)  ,
 \nn\\
 \Smn{c0}0 & = \frac{2\ii}{3} , \qquad
 \Smn{c0}1 = \frac{2\ii}{3} ,\qquad
 \Smn{cc}0 = \frac{6}{\xs^2 x_0^3}, \qquad
 \Smn{cc}1 = \frac{135}{\xs^2 x_0^5},
 \nn\\
 \Smn{cs}0 &= \frac{1}{24\xs^4 x_0} \Big[\big(-\xs^7 + \xs^6 - 14\xs^5 + 18\xs^4 - 48\xs^3 - 96\xs^2 - 144\xs - 144\big)\ee^{-\xs} + E_1(\xs)\xs^6(\xs^2 + 12) \Big] ,
 \nn\\
 \Smn{cs}1 &= -\frac{3\ii}{4\xs^5 x_0^2} \Big[\big(-\xs^7 + \xs^6 - 2\xs^5 + 6\xs^4 + 48\xs^3 + 168\xs^2 + 360\xs + 360\big)\ee^{-\xs} + E_1(\xs)\xs^8 \Big],
 \nn\\
 \Smn{s0}0 &= 0 , \qquad
 \Smn{s0}1 = x_0^2\frac{6\ii }{5\xs^2}\ee^{\ii \xs} , \qquad
 \Smn{sc}0 = 0 ,
 \nn\\
 \Smn{sc}1 &= \frac{1}{32\xs^4 x_0^3} \Big[ \big(\xs^{10} + 18\xs^8\big)E_1( -\ii\xs) - \Big(\xs^9\ii + 16\ii\xs^7 + \xs^8 - 12\ii\xs^5 + 12\xs^6 - 288\ii\xs^3 + 12\xs^4 + 4320\ii\xs \nn\\
 &\quad + 1728\xs^2 - 4320 \Big)\ee^{\ii\xs}\Big] , \qquad
 \Smn{ss}0 = 0 ,
 \nn\\
 \Smn{ss}1 &= -\frac{\ii}{2}\xs E_1(\xs-\ii\xs) \big(\xs^2 + 9\big)+ \frac{1}{\xs^7}\ee^{(-1+\ii)\xs}\bigg[\frac{1}{4}(-1 + \ii)\xs^9 + \frac{1}{4}\xs^8 + \frac{1}{2}(-5 + 4\ii)\xs^7 + \frac{1}{4}(9 + 3\ii)\xs^6 + \frac{1}{4}(9 + 57\ii)\xs^5 \nn\\
 & \quad + \frac{1}{4}(-72 + 177\ii)\xs^4 + (-108 + 72\ii)\xs^3 - (270 + 18\ii)\xs^2 - 270(1 + \ii)\xs -270\ii \bigg].
 \eal
Here, we have used the exponential integral function defined as
$\dpst E_1(x)= \int_1^\infty \xi^{-1} \ee^{-x \xi} \: \dm \xi$.

\end{widetext}


\begin{thebibliography}{34}%
\makeatletter
\providecommand \@ifxundefined [1]{%
 \@ifx{#1\undefined}
}%
\providecommand \@ifnum [1]{%
 \ifnum #1\expandafter \@firstoftwo
 \else \expandafter \@secondoftwo
 \fi
}%
\providecommand \@ifx [1]{%
 \ifx #1\expandafter \@firstoftwo
 \else \expandafter \@secondoftwo
 \fi
}%
\providecommand \natexlab [1]{#1}%
\providecommand \enquote  [1]{``#1''}%
\providecommand \bibnamefont  [1]{#1}%
\providecommand \bibfnamefont [1]{#1}%
\providecommand \citenamefont [1]{#1}%
\providecommand \href@noop [0]{\@secondoftwo}%
\providecommand \href [0]{\begingroup \@sanitize@url \@href}%
\providecommand \@href[1]{\@@startlink{#1}\@@href}%
\providecommand \@@href[1]{\endgroup#1\@@endlink}%
\providecommand \@sanitize@url [0]{\catcode `\\12\catcode `\$12\catcode
  `\&12\catcode `\#12\catcode `\^12\catcode `\_12\catcode `\%12\relax}%
\providecommand \@@startlink[1]{}%
\providecommand \@@endlink[0]{}%
\providecommand \url  [0]{\begingroup\@sanitize@url \@url }%
\providecommand \@url [1]{\endgroup\@href {#1}{\urlprefix }}%
\providecommand \urlprefix  [0]{URL }%
\providecommand \Eprint [0]{\href }%
\providecommand \doibase [0]{https://doi.org/}%
\providecommand \selectlanguage [0]{\@gobble}%
\providecommand \bibinfo  [0]{\@secondoftwo}%
\providecommand \bibfield  [0]{\@secondoftwo}%
\providecommand \translation [1]{[#1]}%
\providecommand \BibitemOpen [0]{}%
\providecommand \bibitemStop [0]{}%
\providecommand \bibitemNoStop [0]{.\EOS\space}%
\providecommand \EOS [0]{\spacefactor3000\relax}%
\providecommand \BibitemShut  [1]{\csname bibitem#1\endcsname}%
\let\auto@bib@innerbib\@empty
\bibitem [{\citenamefont {King}(1934)}]{King1934}%
  \BibitemOpen
  \bibfield  {author} {\bibinfo {author} {\bibfnamefont {L.~V.}\ \bibnamefont
  {King}},\ }\bibfield  {title} {\bibinfo {title} {On the acoustic radiation
  pressure on spheres},\ }\href {https://doi.org/10.1098/rspa.1934.0215}
  {\bibfield  {journal} {\bibinfo  {journal} {Proc. R. Soc. London, Ser. A}\
  }\textbf {\bibinfo {volume} {147}},\ \bibinfo {pages} {212} (\bibinfo {year}
  {1934})}\BibitemShut {NoStop}%
\bibitem [{\citenamefont {Yosioka}\ and\ \citenamefont
  {Kawasima}(1955)}]{Yosioka1955}%
  \BibitemOpen
  \bibfield  {author} {\bibinfo {author} {\bibfnamefont {K.}~\bibnamefont
  {Yosioka}}\ and\ \bibinfo {author} {\bibfnamefont {Y.}~\bibnamefont
  {Kawasima}},\ }\bibfield  {title} {\bibinfo {title} {Acoustic radiation
  pressure on a compressible sphere},\ }\href@noop {} {\bibfield  {journal}
  {\bibinfo  {journal} {Acustica}\ }\textbf {\bibinfo {volume} {5}},\ \bibinfo
  {pages} {167} (\bibinfo {year} {1955})}\BibitemShut {NoStop}%
\bibitem [{\citenamefont {Gorkov}(1962)}]{Gorkov1962}%
  \BibitemOpen
  \bibfield  {author} {\bibinfo {author} {\bibfnamefont {L.~P.}\ \bibnamefont
  {Gorkov}},\ }\bibfield  {title} {\bibinfo {title} {On the forces acting on a
  small particle in an acoustical field in an ideal fluid},\ }\href@noop {}
  {\bibfield  {journal} {\bibinfo  {journal} {Sov. Phys.--Dokl.}\ }\textbf
  {\bibinfo {volume} {6}},\ \bibinfo {pages} {773} (\bibinfo {year} {1962})},\
  \bibinfo {note} {[Doklady Akademii Nauk SSSR \textbf{140}, 88
  (1961)]}\BibitemShut {NoStop}%
\bibitem [{\citenamefont {Doinikov}(1994{\natexlab{a}})}]{Doinikov1994a}%
  \BibitemOpen
  \bibfield  {author} {\bibinfo {author} {\bibfnamefont {A.}~\bibnamefont
  {Doinikov}},\ }\bibfield  {title} {\bibinfo {title} {Acoustic radiation
  pressure on a rigid sphere in a viscous fluid},\ }\href
  {https://doi.org/10.1098/rspa.1994.0150} {\bibfield  {journal} {\bibinfo
  {journal} {Proc. R. Soc. A: Math. Phys. Eng. Sci.}\ }\textbf {\bibinfo
  {volume} {447}},\ \bibinfo {pages} {447} (\bibinfo {year}
  {1994}{\natexlab{a}})}\BibitemShut {NoStop}%
\bibitem [{\citenamefont {Doinikov}(1994{\natexlab{b}})}]{Doinikov1994}%
  \BibitemOpen
  \bibfield  {author} {\bibinfo {author} {\bibfnamefont {A.~A.}\ \bibnamefont
  {Doinikov}},\ }\bibfield  {title} {\bibinfo {title} {Acoustic radiation
  pressure on a compressible sphere in a viscous fluid},\ }\href
  {https://doi.org/10.1017/S0022112094001096} {\bibfield  {journal} {\bibinfo
  {journal} {J. Fluid Mech.}\ }\textbf {\bibinfo {volume} {267}},\ \bibinfo
  {pages} {1} (\bibinfo {year} {1994}{\natexlab{b}})}\BibitemShut {NoStop}%
\bibitem [{\citenamefont {Doinikov}(1997{\natexlab{a}})}]{Doinikov1997}%
  \BibitemOpen
  \bibfield  {author} {\bibinfo {author} {\bibfnamefont {A.~A.}\ \bibnamefont
  {Doinikov}},\ }\bibfield  {title} {\bibinfo {title} {Acoustic radiation force
  on a spherical particle in a viscous heat-conducting fluid .1. general
  formula},\ }\href {https://doi.org/10.1121/1.418035} {\bibfield  {journal}
  {\bibinfo  {journal} {J. Acoust. Soc. Am.}\ }\textbf {\bibinfo {volume}
  {101}},\ \bibinfo {pages} {713} (\bibinfo {year}
  {1997}{\natexlab{a}})}\BibitemShut {NoStop}%
\bibitem [{\citenamefont {Doinikov}(1997{\natexlab{b}})}]{Doinikov1997a}%
  \BibitemOpen
  \bibfield  {author} {\bibinfo {author} {\bibfnamefont {A.~A.}\ \bibnamefont
  {Doinikov}},\ }\bibfield  {title} {\bibinfo {title} {Acoustic radiation force
  on a spherical particle in a viscous heat-conducting fluid .2. force on a
  rigid sphere},\ }\href {https://doi.org/10.1121/1.418036} {\bibfield
  {journal} {\bibinfo  {journal} {J. Acoust. Soc. Am.}\ }\textbf {\bibinfo
  {volume} {101}},\ \bibinfo {pages} {722} (\bibinfo {year}
  {1997}{\natexlab{b}})}\BibitemShut {NoStop}%
\bibitem [{\citenamefont {Doinikov}(1997{\natexlab{c}})}]{Doinikov1997b}%
  \BibitemOpen
  \bibfield  {author} {\bibinfo {author} {\bibfnamefont {A.~A.}\ \bibnamefont
  {Doinikov}},\ }\bibfield  {title} {\bibinfo {title} {{Acoustic radiation
  force on a spherical particle in a viscous heat-conducting fluid. 3. Force on
  a liquid drop}},\ }\href {https://doi.org/10.1121/1.417961} {\bibfield
  {journal} {\bibinfo  {journal} {J. Acoust. Soc. Am.}\ }\textbf {\bibinfo
  {volume} {101}},\ \bibinfo {pages} {731} (\bibinfo {year}
  {1997}{\natexlab{c}})}\BibitemShut {NoStop}%
\bibitem [{\citenamefont {Settnes}\ and\ \citenamefont
  {Bruus}(2012)}]{Settnes2012}%
  \BibitemOpen
  \bibfield  {author} {\bibinfo {author} {\bibfnamefont {M.}~\bibnamefont
  {Settnes}}\ and\ \bibinfo {author} {\bibfnamefont {H.}~\bibnamefont
  {Bruus}},\ }\bibfield  {title} {\bibinfo {title} {Forces acting on a small
  particle in an acoustical field in a viscous fluid},\ }\href
  {https://doi.org/10.1103/PhysRevE.85.016327} {\bibfield  {journal} {\bibinfo
  {journal} {Phys. Rev. E}\ }\textbf {\bibinfo {volume} {85}},\ \bibinfo
  {pages} {016327} (\bibinfo {year} {2012})}\BibitemShut {NoStop}%
\bibitem [{\citenamefont {Karlsen}\ and\ \citenamefont
  {Bruus}(2015)}]{Karlsen2015}%
  \BibitemOpen
  \bibfield  {author} {\bibinfo {author} {\bibfnamefont {J.~T.}\ \bibnamefont
  {Karlsen}}\ and\ \bibinfo {author} {\bibfnamefont {H.}~\bibnamefont
  {Bruus}},\ }\bibfield  {title} {\bibinfo {title} {Forces acting on a small
  particle in an acoustical field in a thermoviscous fluid},\ }\href
  {https://doi.org/10.1103/PhysRevE.92.043010} {\bibfield  {journal} {\bibinfo
  {journal} {Phys. Rev. E}\ }\textbf {\bibinfo {volume} {92}},\ \bibinfo
  {pages} {043010} (\bibinfo {year} {2015})}\BibitemShut {NoStop}%
\bibitem [{\citenamefont {Doinikov}\ \emph {et~al.}(2021)\citenamefont
  {Doinikov}, \citenamefont {Fankhauser},\ and\ \citenamefont
  {Dual}}]{Doinikov2021}%
  \BibitemOpen
  \bibfield  {author} {\bibinfo {author} {\bibfnamefont {A.~A.}\ \bibnamefont
  {Doinikov}}, \bibinfo {author} {\bibfnamefont {J.}~\bibnamefont
  {Fankhauser}},\ and\ \bibinfo {author} {\bibfnamefont {J.}~\bibnamefont
  {Dual}},\ }\bibfield  {title} {\bibinfo {title} {Nonlinear dynamics of a
  solid particle in an acoustically excited viscoelastic fluid. {I. Acoustic}
  streaming},\ }\href {https://doi.org/10.1103/PhysRevE.104.065107} {\bibfield
  {journal} {\bibinfo  {journal} {Phys. Rev. E}\ }\textbf {\bibinfo {volume}
  {104}},\ \bibinfo {pages} {065107} (\bibinfo {year} {2021})}\BibitemShut
  {NoStop}%
\bibitem [{\citenamefont {Winckelmann}\ and\ \citenamefont
  {Bruus}(2023)}]{Winckelmann2023}%
  \BibitemOpen
  \bibfield  {author} {\bibinfo {author} {\bibfnamefont {B.~G.}\ \bibnamefont
  {Winckelmann}}\ and\ \bibinfo {author} {\bibfnamefont {H.}~\bibnamefont
  {Bruus}},\ }\bibfield  {title} {\bibinfo {title} {Acoustic radiation force on
  a spherical thermoviscous particle in a thermoviscous fluid including
  scattering and microstreaming},\ }\href
  {https://doi.org/10.1103/PhysRevE.107.065103} {\bibfield  {journal} {\bibinfo
   {journal} {Phys. Rev. E}\ }\textbf {\bibinfo {volume} {107}},\ \bibinfo
  {pages} {065103} (\bibinfo {year} {2023})}\BibitemShut {NoStop}%
\bibitem [{\citenamefont {Joergensen}\ and\ \citenamefont
  {Bruus}(2021)}]{Joergensen2021}%
  \BibitemOpen
  \bibfield  {author} {\bibinfo {author} {\bibfnamefont {J.~H.}\ \bibnamefont
  {Joergensen}}\ and\ \bibinfo {author} {\bibfnamefont {H.}~\bibnamefont
  {Bruus}},\ }\bibfield  {title} {\bibinfo {title} {Theory of pressure
  acoustics with thermoviscous boundary layers and streaming in elastic
  cavities},\ }\href {https://doi.org/10.1121/10.0005005} {\bibfield  {journal}
  {\bibinfo  {journal} {J. Acoust. Soc. Am.}\ }\textbf {\bibinfo {volume}
  {149}},\ \bibinfo {pages} {3599} (\bibinfo {year} {2021})}\BibitemShut
  {NoStop}%
\bibitem [{\citenamefont {Lee}\ and\ \citenamefont {Wang}(1984)}]{Lee1984}%
  \BibitemOpen
  \bibfield  {author} {\bibinfo {author} {\bibfnamefont {C.~P.}\ \bibnamefont
  {Lee}}\ and\ \bibinfo {author} {\bibfnamefont {T.~G.}\ \bibnamefont {Wang}},\
  }\bibfield  {title} {\bibinfo {title} {The acoustic radiation force on a
  heated (or cooled) rigid sphere--theory},\ }\href
  {https://doi.org/10.1121/1.390304} {\bibfield  {journal} {\bibinfo  {journal}
  {J. Acoust. Soc. Am.}\ }\textbf {\bibinfo {volume} {75}},\ \bibinfo {pages}
  {88} (\bibinfo {year} {1984})}\BibitemShut {NoStop}%
\bibitem [{\citenamefont {Lee}\ and\ \citenamefont {Wang}(1988)}]{Lee1988}%
  \BibitemOpen
  \bibfield  {author} {\bibinfo {author} {\bibfnamefont {C.~P.}\ \bibnamefont
  {Lee}}\ and\ \bibinfo {author} {\bibfnamefont {T.~G.}\ \bibnamefont {Wang}},\
  }\bibfield  {title} {\bibinfo {title} {Acoustic radiation force on a heated
  sphere including effects of heat transfer and acoustic streaming},\ }\href
  {https://doi.org/10.1121/1.395936} {\bibfield  {journal} {\bibinfo  {journal}
  {J. Acoust. Soc. Am.}\ }\textbf {\bibinfo {volume} {83}},\ \bibinfo {pages}
  {1324} (\bibinfo {year} {1988})}\BibitemShut {NoStop}%
\bibitem [{Note1()}]{Note1}%
  \BibitemOpen
  \bibinfo {note} {See Supplemental Material at \protect \url
  {https://bruus-lab.dk/files/Winckelmann_Frad_heated_sphere_suppl.zip} for
  details on numerical simulations in COMSOL Multiphysics and comments on
  temperature dependent material parameters.}\BibitemShut {Stop}%
\bibitem [{\citenamefont {Goldenberg}\ and\ \citenamefont
  {Tranter}(1952)}]{Goldenberg1952}%
  \BibitemOpen
  \bibfield  {author} {\bibinfo {author} {\bibfnamefont {H.}~\bibnamefont
  {Goldenberg}}\ and\ \bibinfo {author} {\bibfnamefont {C.~J.}\ \bibnamefont
  {Tranter}},\ }\bibfield  {title} {\bibinfo {title} {Heat flow in an infinite
  medium heated by a sphere},\ }\href
  {https://doi.org/10.1088/0508-3443/3/9/307} {\bibfield  {journal} {\bibinfo
  {journal} {Br. J. Appl. Phys.}\ }\textbf {\bibinfo {volume} {3}},\ \bibinfo
  {pages} {296} (\bibinfo {year} {1952})}\BibitemShut {NoStop}%
\bibitem [{\citenamefont {Muller}\ and\ \citenamefont
  {Bruus}(2014)}]{Muller2014}%
  \BibitemOpen
  \bibfield  {author} {\bibinfo {author} {\bibfnamefont {P.~B.}\ \bibnamefont
  {Muller}}\ and\ \bibinfo {author} {\bibfnamefont {H.}~\bibnamefont {Bruus}},\
  }\bibfield  {title} {\bibinfo {title} {Numerical study of thermoviscous
  effects in ultrasound-induced acoustic streaming in microchannels},\ }\href
  {https://doi.org/10.1103/PhysRevE.90.043016} {\bibfield  {journal} {\bibinfo
  {journal} {Phys. Rev. E}\ }\textbf {\bibinfo {volume} {90}},\ \bibinfo
  {pages} {043016} (\bibinfo {year} {2014})}\BibitemShut {NoStop}%
\bibitem [{\citenamefont {Wagner}\ and\ \citenamefont
  {Pruss}(2002)}]{Wagner2002}%
  \BibitemOpen
  \bibfield  {author} {\bibinfo {author} {\bibfnamefont {W.}~\bibnamefont
  {Wagner}}\ and\ \bibinfo {author} {\bibfnamefont {A.}~\bibnamefont {Pruss}},\
  }\bibfield  {title} {\bibinfo {title} {The iapws formulation 1995 for the
  thermodynamic properties of ordinary water substance for general and
  scientific use},\ }\href {https://doi.org/10.1063/1.1461829} {\bibfield
  {journal} {\bibinfo  {journal} {J. Phys. Chem. Ref. Data}\ }\textbf {\bibinfo
  {volume} {31}},\ \bibinfo {pages} {387} (\bibinfo {year} {2002})}\BibitemShut
  {NoStop}%
\bibitem [{\citenamefont {Huber}\ \emph {et~al.}(2009)\citenamefont {Huber},
  \citenamefont {Perkins}, \citenamefont {Laesecke}, \citenamefont {Friend},
  \citenamefont {Sengers}, \citenamefont {Assael}, \citenamefont {Metaxa},
  \citenamefont {Vogel}, \citenamefont {Mares},\ and\ \citenamefont
  {Miyagawa}}]{Huber2009}%
  \BibitemOpen
  \bibfield  {author} {\bibinfo {author} {\bibfnamefont {M.~L.}\ \bibnamefont
  {Huber}}, \bibinfo {author} {\bibfnamefont {R.~A.}\ \bibnamefont {Perkins}},
  \bibinfo {author} {\bibfnamefont {A.}~\bibnamefont {Laesecke}}, \bibinfo
  {author} {\bibfnamefont {D.~G.}\ \bibnamefont {Friend}}, \bibinfo {author}
  {\bibfnamefont {J.~V.}\ \bibnamefont {Sengers}}, \bibinfo {author}
  {\bibfnamefont {M.~J.}\ \bibnamefont {Assael}}, \bibinfo {author}
  {\bibfnamefont {I.~N.}\ \bibnamefont {Metaxa}}, \bibinfo {author}
  {\bibfnamefont {E.}~\bibnamefont {Vogel}}, \bibinfo {author} {\bibfnamefont
  {R.}~\bibnamefont {Mares}},\ and\ \bibinfo {author} {\bibfnamefont
  {K.}~\bibnamefont {Miyagawa}},\ }\bibfield  {title} {\bibinfo {title} {New
  international formulation for the viscosity of h2o},\ }\href
  {https://doi.org/10.1063/1.3088050} {\bibfield  {journal} {\bibinfo
  {journal} {J. Phys. Chem. Ref. Data}\ }\textbf {\bibinfo {volume} {38}},\
  \bibinfo {pages} {101} (\bibinfo {year} {2009})}\BibitemShut {NoStop}%
\bibitem [{\citenamefont {Huber}\ \emph {et~al.}(2012)\citenamefont {Huber},
  \citenamefont {Perkins}, \citenamefont {Friend}, \citenamefont {Sengers},
  \citenamefont {Assael}, \citenamefont {Metaxa}, \citenamefont {Miyagawa},
  \citenamefont {Hellmann},\ and\ \citenamefont {Vogel}}]{Huber2012}%
  \BibitemOpen
  \bibfield  {author} {\bibinfo {author} {\bibfnamefont {M.~L.}\ \bibnamefont
  {Huber}}, \bibinfo {author} {\bibfnamefont {R.~A.}\ \bibnamefont {Perkins}},
  \bibinfo {author} {\bibfnamefont {D.~G.}\ \bibnamefont {Friend}}, \bibinfo
  {author} {\bibfnamefont {J.~V.}\ \bibnamefont {Sengers}}, \bibinfo {author}
  {\bibfnamefont {M.~J.}\ \bibnamefont {Assael}}, \bibinfo {author}
  {\bibfnamefont {I.~N.}\ \bibnamefont {Metaxa}}, \bibinfo {author}
  {\bibfnamefont {K.}~\bibnamefont {Miyagawa}}, \bibinfo {author}
  {\bibfnamefont {R.}~\bibnamefont {Hellmann}},\ and\ \bibinfo {author}
  {\bibfnamefont {E.}~\bibnamefont {Vogel}},\ }\bibfield  {title} {\bibinfo
  {title} {New international formulation for the thermal conductivity of h2o},\
  }\href {https://doi.org/10.1063/1.4738955} {\bibfield  {journal} {\bibinfo
  {journal} {J. Phys. Chem. Ref. Data}\ }\textbf {\bibinfo {volume} {41}},\
  \bibinfo {pages} {033102} (\bibinfo {year} {2012})}\BibitemShut {NoStop}%
\bibitem [{\citenamefont {Coupland}\ and\ \citenamefont
  {McClements}(1997)}]{Coupland1997}%
  \BibitemOpen
  \bibfield  {author} {\bibinfo {author} {\bibfnamefont {J.~N.}\ \bibnamefont
  {Coupland}}\ and\ \bibinfo {author} {\bibfnamefont {D.~J.}\ \bibnamefont
  {McClements}},\ }\bibfield  {title} {\bibinfo {title} {Physical properties of
  liquid edible oils},\ }\href {https://doi.org/10.1007/s11746-997-0077-1}
  {\bibfield  {journal} {\bibinfo  {journal} {J. Am. Oil Chem. Soc.}\ }\textbf
  {\bibinfo {volume} {74}},\ \bibinfo {pages} {1559} (\bibinfo {year}
  {1997})}\BibitemShut {NoStop}%
\bibitem [{\citenamefont {Noureddini}\ \emph {et~al.}(1992)\citenamefont
  {Noureddini}, \citenamefont {Teoh},\ and\ \citenamefont
  {Davis~Clements}}]{Noureddini1992}%
  \BibitemOpen
  \bibfield  {author} {\bibinfo {author} {\bibfnamefont {H.}~\bibnamefont
  {Noureddini}}, \bibinfo {author} {\bibfnamefont {B.}~\bibnamefont {Teoh}},\
  and\ \bibinfo {author} {\bibfnamefont {L.}~\bibnamefont {Davis~Clements}},\
  }\bibfield  {title} {\bibinfo {title} {Viscosities of vegetable oils and
  fatty acids},\ }\href {https://doi.org/10.1007/BF02637678} {\bibfield
  {journal} {\bibinfo  {journal} {J. Am. Oil Chem. Soc.}\ }\textbf {\bibinfo
  {volume} {69}},\ \bibinfo {pages} {1189} (\bibinfo {year}
  {1992})}\BibitemShut {NoStop}%
\bibitem [{\citenamefont {Ghosh}\ \emph {et~al.}(2017)\citenamefont {Ghosh},
  \citenamefont {Holmes},\ and\ \citenamefont {Povey}}]{Ghosh2017}%
  \BibitemOpen
  \bibfield  {author} {\bibinfo {author} {\bibfnamefont {S.}~\bibnamefont
  {Ghosh}}, \bibinfo {author} {\bibfnamefont {M.}~\bibnamefont {Holmes}},\ and\
  \bibinfo {author} {\bibfnamefont {M.}~\bibnamefont {Povey}},\ }\bibfield
  {title} {\bibinfo {title} {Temperature dependence of bulk viscosity in edible
  oils using acoustic spectroscopy},\ }\href
  {https://doi.org/10.4172/2157-7110.1000676} {\bibfield  {journal} {\bibinfo
  {journal} {J. Food Process. Technol.}\ }\textbf {\bibinfo {volume} {8}},\
  \bibinfo {pages} {1000676} (\bibinfo {year} {2017})}\BibitemShut {NoStop}%
\bibitem [{\citenamefont {Sun}\ \emph {et~al.}(1988)\citenamefont {Sun},
  \citenamefont {Schouten}, \citenamefont {Trappeniers},\ and\ \citenamefont
  {Biswas}}]{Sun1988}%
  \BibitemOpen
  \bibfield  {author} {\bibinfo {author} {\bibfnamefont {T.}~\bibnamefont
  {Sun}}, \bibinfo {author} {\bibfnamefont {J.}~\bibnamefont {Schouten}},
  \bibinfo {author} {\bibfnamefont {N.}~\bibnamefont {Trappeniers}},\ and\
  \bibinfo {author} {\bibfnamefont {S.}~\bibnamefont {Biswas}},\ }\bibfield
  {title} {\bibinfo {title} {Measurements of the densities of liquid benzene,
  cyclohexane, methanol, and ethanol as functions of temperature at 0.1 mpa},\
  }\href {https://doi.org/https://doi.org/10.1016/0021-9614(88)90115-2}
  {\bibfield  {journal} {\bibinfo  {journal} {J. Chem. Thermodyn.}\ }\textbf
  {\bibinfo {volume} {20}},\ \bibinfo {pages} {1089} (\bibinfo {year}
  {1988})}\BibitemShut {NoStop}%
\bibitem [{\citenamefont {Dukhin}\ and\ \citenamefont
  {Goetz}(2009)}]{Dukhin2009}%
  \BibitemOpen
  \bibfield  {author} {\bibinfo {author} {\bibfnamefont {A.~S.}\ \bibnamefont
  {Dukhin}}\ and\ \bibinfo {author} {\bibfnamefont {P.~J.}\ \bibnamefont
  {Goetz}},\ }\bibfield  {title} {\bibinfo {title} {Bulk viscosity and
  compressibility measurement using acoustic spectroscopy},\ }\href
  {https://doi.org/10.1063/1.3095471} {\bibfield  {journal} {\bibinfo
  {journal} {J Chem Phys}\ }\textbf {\bibinfo {volume} {130}},\ \bibinfo
  {pages} {124519} (\bibinfo {year} {2009})}\BibitemShut {NoStop}%
\bibitem [{Che()}]{Cheric_ethanol}%
  \BibitemOpen
  \href@noop {} {\emph {\bibinfo {title} {Pure Component Properties}}},\
  \bibinfo {organization} {CHERIC, Chemical Engineering and Materials Research
  Incformation Center},\ \bibinfo {note}
  {\url{https://www.cheric.org/research/kdb/hcprop/showprop.php?cmpid=818},
  accessed 13 May 2023}\BibitemShut {NoStop}%
\bibitem [{Ond()}]{OndaCorp}%
  \BibitemOpen
  \href@noop {} {\emph {\bibinfo {title} {Tables of Acoustic Properties of
  Materials: Plastics}}},\ \bibinfo {organization} {Onda Corporation},\
  \bibinfo {note}
  {\url{https://www.ondacorp.com/wp-content/uploads/2020/09/Plastics.pdf},
  accessed 13 May 2023}\BibitemShut {NoStop}%
\bibitem [{\citenamefont {Ellis}\ and\ \citenamefont
  {Smith}(2008)}]{CRCpolymers}%
  \BibitemOpen
  \bibinfo {editor} {\bibfnamefont {B.}~\bibnamefont {Ellis}}\ and\ \bibinfo
  {editor} {\bibfnamefont {R.}~\bibnamefont {Smith}},\ eds.,\ \href
  {https://doi.org/https://doi.org/10.1201/9781420005707} {\emph {\bibinfo
  {title} {Polymers: A Properties Database}}},\ \bibinfo {edition} {2nd}\ ed.\
  (\bibinfo  {publisher} {CRC Press},\ \bibinfo {address} {Boca Raton, FL},\
  \bibinfo {year} {2008})\BibitemShut {NoStop}%
\bibitem [{\citenamefont {Domalski}\ and\ \citenamefont
  {Hearing}(1996)}]{Domalski1996}%
  \BibitemOpen
  \bibfield  {author} {\bibinfo {author} {\bibfnamefont {E.~S.}\ \bibnamefont
  {Domalski}}\ and\ \bibinfo {author} {\bibfnamefont {E.~D.}\ \bibnamefont
  {Hearing}},\ }\bibfield  {title} {\bibinfo {title} {{Heat Capacities and
  Entropies of Organic Compounds in the Condensed Phase. Volume III}},\ }\href
  {https://doi.org/10.1063/1.555985} {\bibfield  {journal} {\bibinfo  {journal}
  {J. Phys. Chem. Ref. Data}\ }\textbf {\bibinfo {volume} {25}},\ \bibinfo
  {pages} {1} (\bibinfo {year} {1996})}\BibitemShut {NoStop}%
\bibitem [{\citenamefont {Chang}\ and\ \citenamefont
  {Bestul}(1968)}]{Chang1968}%
  \BibitemOpen
  \bibfield  {author} {\bibinfo {author} {\bibfnamefont {S.~S.}\ \bibnamefont
  {Chang}}\ and\ \bibinfo {author} {\bibfnamefont {A.~B.}\ \bibnamefont
  {Bestul}},\ }\bibfield  {title} {\bibinfo {title} {{Heat capacities for
  atactic polystyrene of narrow molecular weight distribution to 360~K.}},\
  }\href {https://doi.org/10.1002/pol.1968.160060505} {\bibfield  {journal}
  {\bibinfo  {journal} {J. Polym. Sci. A-2 Polym. Phys.}\ }\textbf {\bibinfo
  {volume} {6}},\ \bibinfo {pages} {849} (\bibinfo {year} {1968})}\BibitemShut
  {NoStop}%
\bibitem [{\citenamefont {Smith}\ and\ \citenamefont
  {Wiggins}(1972)}]{Smith1972}%
  \BibitemOpen
  \bibfield  {author} {\bibinfo {author} {\bibfnamefont {D.~M.}\ \bibnamefont
  {Smith}}\ and\ \bibinfo {author} {\bibfnamefont {T.~A.}\ \bibnamefont
  {Wiggins}},\ }\bibfield  {title} {\bibinfo {title} {Sound speeds and laser
  induced damage in polystyrene},\ }\href
  {https://doi.org/10.1364/AO.11.002680} {\bibfield  {journal} {\bibinfo
  {journal} {Appl. Opt.}\ }\textbf {\bibinfo {volume} {11}},\ \bibinfo {pages}
  {2680} (\bibinfo {year} {1972})}\BibitemShut {NoStop}%
\bibitem [{Com(2020)}]{Comsol60}%
  \BibitemOpen
  \href@noop {} {}\bibinfo {organization} {{COMSOL Multiphysics 6.0}} (\bibinfo
  {year} {2020}),\ \bibinfo {note} {\url{http://www.comsol.com}}\BibitemShut
  {NoStop}%
\bibitem [{\citenamefont {Augustsson}\ \emph {et~al.}(2016)\citenamefont
  {Augustsson}, \citenamefont {Karlsen}, \citenamefont {Su}, \citenamefont
  {Bruus},\ and\ \citenamefont {Voldman}}]{Augustsson2016}%
  \BibitemOpen
  \bibfield  {author} {\bibinfo {author} {\bibfnamefont {P.}~\bibnamefont
  {Augustsson}}, \bibinfo {author} {\bibfnamefont {J.~T.}\ \bibnamefont
  {Karlsen}}, \bibinfo {author} {\bibfnamefont {H.-W.}\ \bibnamefont {Su}},
  \bibinfo {author} {\bibfnamefont {H.}~\bibnamefont {Bruus}},\ and\ \bibinfo
  {author} {\bibfnamefont {J.}~\bibnamefont {Voldman}},\ }\bibfield  {title}
  {\bibinfo {title} {Iso-acoustic focusing of cells for size-insensitive
  acousto-mechanical phenotyping},\ }\href
  {https://doi.org/10.1038/ncomms11556} {\bibfield  {journal} {\bibinfo
  {journal} {Nat. Commun.}\ }\textbf {\bibinfo {volume} {7}},\ \bibinfo {pages}
  {11556} (\bibinfo {year} {2016})}\BibitemShut {NoStop}%
\end{thebibliography}

%

\end{document}